\documentclass[times]{article}
\newcommand{\typeof}{0}
 
\usepackage{ifthen}
\newcommand{\condinc}[2]{\ifthenelse{\equal{\typeof}{0}}{#1}{#2}}

\usepackage{latex8}
\usepackage{times}

\usepackage{mathrsfs}

\usepackage{amsmath,amssymb}
\usepackage{xspace}
\usepackage{stmaryrd}
\usepackage{url}

\usepackage{xy}
\xyoption{frame}
\xyoption{arrow}
\xyoption{curve}
\xyoption{arc}
\xyoption{line}
\xyoption{matrix}
\xyoption{tile}
\xyoption{ps}

\usepackage{amsthm}

\theoremstyle{plain}
    \newtheorem{theorem}{Theorem}
    \newtheorem{lemma}{Lemma}
    \newtheorem{corollary}{Corollary}
    \newtheorem{proposition}{Proposition}

    \newtheorem{example}{Example}
\theoremstyle{definition}
    \newtheorem{definition}{Definition}
\theoremstyle{remark}
    \newtheorem{remark}{Remark}
\theoremstyle{plain}

\newcount\mscount
\newdimen\proofrulebreadth \proofrulebreadth=.05em
\newdimen\proofdotseparation \proofdotseparation=1.25ex
\newdimen\proofrulebaseline \proofrulebaseline=2ex
\newcount\proofdotnumber \proofdotnumber=3
\let\then\relax
\def\hfi{\hskip0pt plus.0001fil}
\mathchardef\squigto="3A3B
\newif\ifinsideprooftree\insideprooftreefalse
\newif\ifonleftofproofrule\onleftofproofrulefalse
\newif\ifproofdots\proofdotsfalse
\newif\ifdoubleproof\doubleprooffalse
\let\wereinproofbit\relax
\newdimen\shortenproofleft
\newdimen\shortenproofright
\newdimen\proofbelowshift
\newbox\proofabove
\newbox\proofbelow
\newbox\proofrulename
\def\shiftproofbelow{\let\next\relax\afterassignment\setshiftproofbelow\dimen0 }
\def\shiftproofbelowneg{\def\next{\multiply\dimen0 by-1 }%
\afterassignment\setshiftproofbelow\dimen0 }
\def\setshiftproofbelow{\next\proofbelowshift=\dimen0 }
\def\setproofrulebreadth{\proofrulebreadth}

\def\prooftree{
\ifnum	\lastpenalty=1
\then	\unpenalty
\else	\onleftofproofrulefalse
\fi
\ifonleftofproofrule
\else	\ifinsideprooftree
	\then	\hskip.5em plus1fil
	\fi
\fi
\bgroup
\setbox\proofbelow=\hbox{}\setbox\proofrulename=\hbox{}%
\let\justifies\proofover\let\leadsto\proofoverdots\let\Justifies\proofoverdbl
\let\using\proofusing\let\[\prooftree
\ifinsideprooftree\let\]\endprooftree\fi
\proofdotsfalse\doubleprooffalse
\let\thickness\setproofrulebreadth
\let\shiftright\shiftproofbelow \let\shift\shiftproofbelow
\let\shiftleft\shiftproofbelowneg
\let\ifwasinsideprooftree\ifinsideprooftree
\insideprooftreetrue
\setbox\proofabove=\hbox\bgroup$\displaystyle 
\let\wereinproofbit\prooftree
\shortenproofleft=0pt \shortenproofright=0pt \proofbelowshift=0pt
\onleftofproofruletrue\penalty1
}

\def\eproofbit{
\ifx	\wereinproofbit\prooftree
\then	\ifcase	\lastpenalty
	\then	\shortenproofright=0pt	
	\or	\unpenalty\hfil		
	\or	\unpenalty\unskip	
	\else	\shortenproofright=0pt	
	\fi
\fi
\global\dimen0=\shortenproofleft
\global\dimen1=\shortenproofright
\global\dimen2=\proofrulebreadth
\global\dimen3=\proofbelowshift
\global\dimen4=\proofdotseparation
\global\mscount=\proofdotnumber
$\egroup  
\shortenproofleft=\dimen0
\shortenproofright=\dimen1
\proofrulebreadth=\dimen2
\proofbelowshift=\dimen3
\proofdotseparation=\dimen4
\proofdotnumber=\mscount
}

\def\proofover{
\eproofbit 
\setbox\proofbelow=\hbox\bgroup 
\let\wereinproofbit\proofover
$\displaystyle
}%
\def\proofoverdbl{
\eproofbit 
\doubleprooftrue
\setbox\proofbelow=\hbox\bgroup 
\let\wereinproofbit\proofoverdbl
$\displaystyle
}%
\def\proofoverdots{
\eproofbit 
\proofdotstrue
\setbox\proofbelow=\hbox\bgroup 
\let\wereinproofbit\proofoverdots
$\displaystyle
}%
\def\proofusing{
\eproofbit 
\setbox\proofrulename=\hbox\bgroup 
\let\wereinproofbit\proofusing
\kern0.3em$
}

\def\endprooftree{
\eproofbit 
  \dimen5 =0pt
\dimen0=\wd\proofabove \advance\dimen0-\shortenproofleft
\advance\dimen0-\shortenproofright
\dimen1=.5\dimen0 \advance\dimen1-.5\wd\proofbelow
\dimen4=\dimen1
\advance\dimen1\proofbelowshift \advance\dimen4-\proofbelowshift
\ifdim	\dimen1<0pt
\then	\advance\shortenproofleft\dimen1
	\advance\dimen0-\dimen1
	\dimen1=0pt
	\ifdim  \shortenproofleft<0pt
        \then   \setbox\proofabove=\hbox{%
			\kern-\shortenproofleft\unhbox\proofabove}%
                \shortenproofleft=0pt
        \fi
\fi
\ifdim	\dimen4<0pt
\then	\advance\shortenproofright\dimen4
	\advance\dimen0-\dimen4
	\dimen4=0pt
\fi
\ifdim	\shortenproofright<\wd\proofrulename
\then	\shortenproofright=\wd\proofrulename
\fi
\dimen2=\shortenproofleft \advance\dimen2 by\dimen1
\dimen3=\shortenproofright\advance\dimen3 by\dimen4
\ifproofdots
\then
	\dimen6=\shortenproofleft \advance\dimen6 .5\dimen0
	\setbox1=\vbox to\proofdotseparation{\vss\hbox{$\cdot$}\vss}
	\setbox0=\hbox{%
		\kern\dimen6
		$\vcenter to\proofdotnumber\proofdotseparation
			{\leaders\box1\vfill}$%
		\unhbox\proofrulename}%
\else	\dimen6=\fontdimen22\the\textfont2 
	\dimen7=\dimen6
	\advance\dimen6by.5\proofrulebreadth
	\advance\dimen7by-.5\proofrulebreadth
	\setbox0=\hbox{%
		\kern\shortenproofleft
		\ifdoubleproof
		\then	\hbox to\dimen0{%
			$\mathsurround0pt\mathord=\mkern-6mu%
			\cleaders\hbox{$\mkern-2mu=\mkern-2mu$}\hfill
			\mkern-6mu\mathord=$}%
		\else	\vrule height\dimen6 depth-\dimen7 width\dimen0
		\fi
		\unhbox\proofrulename}%
	\ht0=\dimen6 \dp0=-\dimen7
\fi
\let\doll\relax
\ifwasinsideprooftree
\then	\let\VBOX\vbox
\else	\ifmmode\else$\let\doll=$\fi
	\let\VBOX\vcenter
\fi
\VBOX	{\baselineskip\proofrulebaseline \lineskip.2ex
	\expandafter\lineskiplimit\ifproofdots0ex\else-0.6ex\fi
	\hbox	spread\dimen5	{\hfi\unhbox\proofabove\hfi}%
	\hbox{\box0}%
	\hbox	{\kern\dimen2 \box\proofbelow}}\doll%
\global\dimen2=\dimen2
\global\dimen3=\dimen3
\egroup 
\ifonleftofproofrule
\then	\shortenproofleft=\dimen2
\fi
\shortenproofright=\dimen3
\onleftofproofrulefalse
\ifinsideprooftree
\then	\hskip.5em plus 1fil \penalty2
\fi
}

\newcommand{\dlinea}{\leavevmode\hrule\vspace{1pt}\hrule\mbox{}}

% ...
\usepackage{mathrsfs}

\mathcode`\<="4268  
\mathcode`\>="5269  

\mathchardef\gt="313E
\mathchardef\lt="313C

\newcommand{\NN}{\mathbb{N}}
\newcommand{\CC}{\mathbb{C}}
\newcommand{\RR}{\mathbb{R}}
\newcommand{\ZZ}{\mathbb{Z}}
\newcommand{\PRR}{\mathbf{P}\mathbb{R}}
\newcommand{\PCC}{\mathbf{P}\mathbb{C}}

\newcommand{\urule}[3]{%
  \prooftree #1 \justifies #2 \using #3 \endprooftree}
\newcommand{\brule}[4]{%
  \prooftree #1\ \ \ #2 \justifies #3 \using #4 \endprooftree}

\newcommand{\lto}[2]{%
  \prooftree #2 \leadsto #1 \endprooftree}
\newcommand{\LT}[3]{%
  \prooftree #2 \leadsto #1 \using #3\endprooftree}

\newcommand{\vseq}[1]{%
\begin{array}{c} #1\end{array}} %

\newcommand{\limp}{\multimap} 

\newcommand{\tupla}[1]{\mathsf{t}_{#1}}
\newcommand{\rapp}{\mathsf{r.a}}
\newcommand{\lapp}{\mathsf{l.a}}
\newcommand{\inlambda}{\mathsf{in.\lambda}}
\newcommand{\Uq}{\mathsf{Uq}}
\newcommand{\nw}{\mathsf{new}}

\newcommand{\lbeta}{\mathsf{l.\beta}}
\newcommand{\qbeta}{\mathsf{q.\beta}}
\newcommand{\cbeta}{\mathsf{c.\beta}}
\newcommand{\lcom}{\mathsf{l.cm}}
\newcommand{\rcom}{\mathsf{r.cm}}

\newcommand{\Conf}{{\cal C}}
\newcommand{\CONF}{\mathsf{C}}
\newcommand{\Ok}[1]{#1 \in \Conf}
\newcommand{\OK}[1]{#1 \in \CONF}
\newcommand{\Qr}{{\cal Q}}
\newcommand{\Vr}{{\cal V}}
\newcommand{\QV}{{\cal QV}}
\newcommand{\Ln}{{\cal L}}
\newcommand{\HS}{\mathcal{H}}
\newcommand{\Hs}[1]{\mathcal{H}(#1)}
\newcommand{\Bs}[1]{\mathcal{B}(#1)}

\newcommand{\Qvt}[1]{\mathbf{Q}(#1)}
\newcommand{\CQ}{\mathbf{\mathsf{CNQ}}}
\newcommand{\Eqt}{\mathsf{EQT}}
\newcommand{\Ncl}{\mathsf{NCL}}
\newcommand{\NF}{\mathsf{NF}}
\newcommand{\dr}[1]{|#1\rangle}

\newcommand{\ncoredto}{\rightarrow_\noncommrul}

\newcommand{\mnseq}[2]{\stackrel{#1,#2}{\longrightarrow} }
\newcommand{\mseq}[1]{\stackrel{#1}{\longrightarrow} }

\newcommand{\nat}[1]{\overline{#1}}
\newcommand{\blist}[1]{[#1]}
\newcommand{\tsuc}{\mathbf{succ}}
\newcommand{\pred}{\mathbf{pred}}
\newcommand{\cons}{\mathbf{cons}}
\newcommand{\sel}{\mathbf{sel}}
\newcommand{\iternat}{\mathbf{iternat}}
\newcommand{\iternataux}{\mathbf{iternataux}}
\newcommand{\iterlist}{\mathbf{iterlist}}
\newcommand{\iterlistaux}{\mathbf{iterlistaux}}
\newcommand{\rec}{\mathbf{rec}}
\newcommand{\recaux}{\mathbf{recaux}}
\newcommand{\append}[1]{\mathbf{append}_{#1}}
\newcommand{\extract}[1]{\mathbf{extract}_{#1}}
\newcommand{\redrul}{\mathscr{L}}
\newcommand{\quantrul}{\mathscr{Q}}
\newcommand{\noncrul}{n\mathscr{C}}
\newcommand{\classrul}{\mathscr{C}}
\newcommand{\commrul}{\mathscr{O}}
\newcommand{\noncommrul}{\mathscr{N}}
\newcommand{\scrul}{\mathscr{M}}
\newcommand{\nqrul}{n\mathscr{Q}}

\newcommand{\ssp}{\;}
\newcommand{\wfj}[1]{#1 \triangleright\,}
\newcommand{\qcalc}{\textsf{Q}-calculus}

\newenvironment{varitemize}
{
\begin{list}{\labelitemi}
{\setlength{\itemsep}{0pt}
 \setlength{\topsep}{0pt}
 \setlength{\parsep}{0pt}
 \setlength{\partopsep}{0pt}
 \setlength{\leftmargin}{15pt}
 \setlength{\rightmargin}{0pt}
 \setlength{\itemindent}{0pt}
 \setlength{\labelsep}{5pt}
 \setlength{\labelwidth}{10pt}
}}
{
 \end{list} 
}

\newcounter{numberone}

\newenvironment{varenumerate}
{
\begin{list}{\arabic{numberone}.}
{
  \usecounter{numberone}
  \setlength{\itemsep}{0pt}
  \setlength{\topsep}{0pt}
  \setlength{\parsep}{0pt}
  \setlength{\partopsep}{0pt}
  \setlength{\leftmargin}{15pt}
  \setlength{\rightmargin}{0pt}
  \setlength{\itemindent}{0pt}
  \setlength{\labelsep}{5pt}
  \setlength{\labelwidth}{15pt}
}}
{
\end{list} 
}

\newcounter{numbertwo}

\newenvironment{varvarenumerate}
{
\begin{list}{\roman{numbertwo}.}
{
  \usecounter{numbertwo}
  \setlength{\itemsep}{0pt}
  \setlength{\topsep}{0pt}
  \setlength{\parsep}{0pt}
  \setlength{\partopsep}{0pt}
  \setlength{\leftmargin}{15pt}
  \setlength{\rightmargin}{0pt}
  \setlength{\itemindent}{0pt}
  \setlength{\labelsep}{5pt}
  \setlength{\labelwidth}{15pt}
}}
{
\end{list} 
}

\newcommand{\new}[1]{\mathtt{new}(#1)}
\newcommand{\pnew}[1]{\mathtt{p\_new}(#1)}
\newcommand{\qnew}[1]{\mathtt{q\_new}(#1)}

\newcommand{\redto}{\to}
\newcommand{\predto}[1]{\to_{#1}}

\newcommand{\rredto}{\stackrel{*}{\redto}}
\newcommand{\rpredto}[1]{\stackrel{*}{\predto{#1}}}

\newdimen\netunit
\newlength\vt\newlength\hz
\newcommand{\lspine}[3]{%
  \save\POS
  @+, s0+<1em,1.5em>@+, s0+/u#3\vt/+/d1.5em/@+, s0+/l#2\hz/@+, 
  s0+/d#3\vt/@+, s0.{s0+/r#2\hz/}+R(.4)@+, s0+/u1.5em/@+,
  s6+<-1em,1.5em>@+,
  {s0 \ar@{-} 's1 's2 's3 's4 's5 's6 's7 s0},
  {s6.s4+C}.s1.s5="SPINE#1",
  s2.s3+C="SPINE#1-AUX",
  @-@-@-@-@-@-@-
  \restore\POS
  }

\newcommand{\rspine}[3]{%
  \save\POS
  @+, s0+<-1em,1.5em>@+, s0+/u#3\vt/+/d1.5em/@+, s0+/r#2\hz/@+, 
  s0+/d#3\vt/@+, s0.{s0+/l#2\hz/}+L(.4)@+, s0+/u1.5em/@+,
  s6+<1em,1.5em>@+,
  {s0 \ar@{-} 's1 's2 's3 's4 's5 's6 's7 s0},
  {s6.s4+C}.s1.s5="SPINE#1",
  s2.s3+C="SPINE#1-AUX",
  @-@-@-@-@-@-@-
  \restore\POS
  }

\title{Quantum Lambda Calculi with Classical Control:\\
       Syntax and Expressive Power }

\author{Ugo Dal Lago\\
{\it  \normalsize Dipartimento di Scienze dell'Informazione}\\
{\it  \normalsize Universit\`a di Bologna}\\ 
{ \normalsize \texttt{dallago@cs.unibo.it}}
\and
Andrea Masini\\
{\it  \normalsize Dipartimento di Informatica}\\
{\it  \normalsize Universit\`a di Verona}\\
{ \normalsize \texttt{andrea.masini@univr.it}}
\and
Margherita Zorzi\\
{\it  \normalsize Dipartimento di Informatica}\\
{\it  \normalsize Universit\`a di Verona}\\
{ \normalsize \texttt{zorzim@sci.univr.it}}
}

\begin{document}

\maketitle

\begin{abstract}
  We study an untyped $\lambda$--calculus with quantum
  data and classical control.  This work stems from
  previous proposals by Selinger and Valiron and by
  Van Tonder. We focus on syntax and expressiveness, 
  rather than (denotational) semantics. We prove subject
  reduction, confluence and a standardization theorem. Moreover,  
  we prove the computational equivalence of the proposed calculus 
  with a suitable class of quantum circuit families.
\end{abstract}

\section{Introduction}


Quantum computing was conceived at the beginning of the eighties, starting from
an idea by Feynman~\cite{Fey82}. It defines an alternative computational
paradigm, based on quantum mechanics~\cite{BasDa05} rather than digital
electronics.  The first proposal for a quantum abstract
computer is due to Deutsch, who introduced quantum
Turing machines~\cite{Deu85}.  Other quantum computational models have
been subsequently defined by Yao (quantum circuits,~\cite{Yao93}) and
Knill (quantum random access machines,~\cite{Knill96}). 


The introduction of quantum abstract machines made it possible to develop a 
complexity theory of quantum computation.
The most remarkable result in quantum complexity theory has been
obtained by Shor, who showed that integers can be factorized in polynomial
time~\cite{Shor94}. Shor's algorithm,
like the majority of quantum algorithmics, is defined as a quantum circuit
family generated by a classical device.

Nowadays, what are the main challenges in quantum computing?
A lot of research is being devoted to understanding whether quantum 
computation can provide efficient algorithms for classically intractable 
problems. In the last years, the impressive results obtained in this
area (e.g. Shor's fast factoring algorithm) have stimulated the 
development of quantum programming languages.  
The situation is not as easy as in the classical case. In addition to 
the concrete technical problems (up to now it is difficult to build even very simple
quantum circuits) there is the necessity of developing adequate
theoretical bases for quantum programming languages ---
even with the best will in the world it is hard to look at quantum
Turing machines as a basis for programming.
\textit{This paper is an attempt to give a contribution to the 
definition of a (higher-order) quantum computational model.}





The first attempt to define a quantum functional programming language
has been done (to our knowledge) in two unpublished papers by 
Maymin~\cite{May96,May97}. Selinger~\cite{Sel04c} rigorously defined
a first-order quantum functional language.  Another interesting
proposal in the framework of first-order quantum functional languages
is the language QML of Altenkirch and Grattage~\cite{AltGra05bis}.




Focusing on higher-order functional programming languages, at least
two distinct proposals have already appeared in the literature: that
by Selinger and Valiron~\cite{SelVal06} and the one by Van
Tonder~\cite{vT04}.  These two approaches seems to go in orthogonal
directions: in the language proposed by Selinger and Valiron data
(registers of qubits) are superimposed while control (lambda terms) is
classical, whereas the approach of Van Tonder is based on the idea of
putting {\it arbitrary $\lambda$--terms in superposition\/}.
But, \textit{is this the right picture}? In order to give an answer
let us examine more closely the two approaches.
\paragraph{Selinger and Valiron's Approach.}
The main goal of the work of Selinger and Valiron is to give
the basis of a typed quantum functional language (with types in
propositional multiplicative and exponential linear logic).
The great merit of Selinger and Valiron is to have
defined a language where only data are superposed, 
and where programs live in a standard classical world. In particular, 
it is not necessary to have ``exotic'' objects such as $\lambda$--terms in
superposition. The approach is well condensed by the slogan: 
\textit{``classical control + quantum data''}.
The proposed calculus, here dubbed $\lambda_{sv}$, 
is based on a call-by-value $\lambda$--calculus enriched with
constants for unitary transformations and an explicit measurement operator.

Unfortunately, the expressive power of $\lambda_{sv}$ has not been
studied yet.  The crucial issue is the following: can we
compare  the expressive power of $\lambda_{sv}$ with the
one of any well known computational model (e.g. quantum Turing machines
or quantum circuits families)?


\paragraph{Van Tonder's Approach.}
The calculus introduced by Van Tonder~\cite{vT04}, called $\lambda_q$,
has the same motivation and a number of \textit{immediate
similarities} with $\lambda_{sv}$, noticeably, the exploitation of
linear types in controlling both copying and erasing of terms.


But there is a glaring difference between $\lambda_q$ and
$\lambda_{sv}$. In fact it
seems that \textit{$\lambda_q$ allows by design arbitrary
superpositions of $\lambda$-terms.}
In our opinion the essence of the approach of Van Tonder is in lemma
5.1 of~\cite{vT04}, where it is stated that ``\textit{two terms $t_1, t_2$ in
superposition differ only for qubits values}''. Moreover, if $t_1$
reduces to $t_1'$ and $t_2$ reduces to $t_2'$, the reduced redex in
$t_1$ is (up to quantum bits) the same redex reduced in $t_2$.
This means $\lambda_q$ has classical control, too: 
it is not possible to superimpose terms differing in a remarkable way, 
i.e. terms with a different computational evolution.

The weak point of Van Tonder's paper, is that some results and proofs are
given too informally.  In particular, the paper 
argues that the proposed calculus is computationally equivalent to quantum 
Turing machines without giving a detailed proof and, more importantly,
without specifying which class of quantum Turing machines is
considered (this is not  pedantry, since there isn't anything like
a Church-Turing thesis in quantum computation~\cite{NiOz02}). But clearly,
such a criticism does not invalidate the foundational importance 
of the approach.

\paragraph{Our Proposal.}
Our goal is to propose an alternative quantum computational paradigm,
proving its computational equivalence with quantum circuits families.

Our work can be seen both as a \textit{continuation} and 
\textit{extension} of the two proposals we have just described. 
\begin{varitemize}
\item
  It is a {\it continuation \/} because we  propose  a
  quantum $\lambda$--calculus with classical control and quantum
  data. We  use a syntax for terms and configurations inspired by
  that of Selinger and Valiron and moreover we implicitly use
  linear logic in a way similar to Van Tonder's $\lambda_q$.
\item
  It is an {\it extension\/} because we have focused on
  the syntactical study of the calculus.  Important classical properties
  such as {\it subject reduction\/} and {\it confluence\/} are given.
  Moreover a novel {\it quantum standardization theorem\/} is
  given.  The {\it expressive
  power\/} of the calculus has been studied in a detailed way
  (to our knowledge, it is the first time
  such a study has been done for a quantum $\lambda$--calculus).
  In order to face the expressive power problem, we  prove the
\textit{
equivalence between our calculus and quantum circuit
  families}.
\end{varitemize}

We have chosen $\lambda$--calculus as a basis of our proposal for a
number of reasons:
\begin{varitemize}
\item 
  first of all, quantum computability and complexity theory are
  quite underdeveloped compared to their classical counterparts; 
  in particular, there is almost no
  result relating classes of (first-order) functions definable in 
  pure and typed $\lambda$--calculi with classes of functions 
  from computability and complexity theory (in contrast with
  classical computability theory~\cite{Kleene37});
\item 
  we hope that our proposal will contribute to the development
  of a ``quantum computationally complete'' functional programming
  language. Quantum Turing machines and quantum circuit families are
  good for computability theory, but quite useless from a programming
  perspective;
\item 
  we believe that the higher--order nature of $\lambda$--calculi could be
  useful for understanding the interactions between the
  classical world (the world of terms) and the quantum world (quantum
  registers).
\end{varitemize}
 The  paper is structured as follows:
\begin{varitemize}
\item
  in Section 2 we give the mathematical background on 
  Hilbert Spaces (in order to define quantum registers);
\item
  in Section 3, a $\lambda$--calculus, called the \qcalc, is
  introduced. The \qcalc\ has classical control and quantum data.
  The calculus is untyped, but is equipped with well-formation
  judgments for terms based on the formulation of linear logic as
  proposed in~\cite{Wad94};
\item
  in Section 4 we syntactically study the \qcalc\ by
  means of a suitable formulation of subject reduction theorem and
  confluence theorems. Noticeably, a configuration is strongly
  normalizing iff it is weakly normalizing;
\item
  in section 5 a further result on the dynamics of the \qcalc\
  is given: for each terminating computation
  there is another ``canonical'', equivalent computation where
  computational steps are performed in the following order:
  \begin{varenumerate}
  \item first, classical reductions: in this phase the quantum register
    is empty and  all the computations steps are classical;
  \item secondly, reductions that build the quantum register;
  \item and finally quantum reductions, applying unitary
    transformations to the quantum register.
  \end{varenumerate}
  Such a property is formally ensured by means of a suitable
  standardization theorem and sheds some further light on 
  the dynamics of computation;
\item
  in Sections 6 we study in detail the equivalence of the
  \qcalc\ with Quantum Circuit Families. The equivalence
  proofs are based on the standardization theorem and on suitable
  encodings.
\end{varitemize}

\section{Mathematical Structures}
This section is devoted to mathematical preliminaries. Clearly, we cannot hope
to be completely self-contained here. See~\cite{NieCh00} for an excellent introduction
to quantum computing.
\subsection{ Quantum Computing Basics}
We informally recall here the basic notations on qubits and quantum registers
(see \cite{NieCh00} for a detailed introduction).
In the next subsection such notations will be (re)defined in a rigorous way.\\
The basic unit of quantum computation is called {\itshape quantum
  bit}, or {\itshape qubit} for short .
The more direct way to represent a quantum bit is by an unitary vector
in the 2-dimensional Hilbert space ${\CC}^{2}$.  Let us denote with
$|0>$ and $|1>$ the elements of an orthonormal basis of $\CC^{2}$.

The states $|0>$ and $|1>$ of a qubit can be seen as the correspondent
states of a classical bit. A qubit, however, can be in other states, different from
$|0>$ and $|1>$.  In fact, every linear combination
$|\psi>=\alpha |0> + \beta |1>$
where $\alpha,\beta\in{\CC}$, and $|\alpha|^{2}+|\beta|^{2}=1$, can be
a possible qubit state. These states are
\emph{superpositions}, and the two values $\alpha$ and $\beta$ are called
\emph{amplitudes}.

While we can determine the state of a classical bit, for a qubit we
can't establish with the same precision what is it's quantum state,
namely the values of $\alpha$ and $\beta$:
quantum mechanics says that a measurement of a qubit with state $\alpha |0> + \beta
|1>$ has the effect of changing the state of the qubit to $|0>$ with probability
$|\alpha|^{2}$ and to $|1>$ with probability $|\beta|^{2}$.

 
In computational models, we need a generalization
of the notion of a qubit, namely the so called \textit{quantum register}
\cite{NiOz02, Sel04c, SelVal06, vT04}.
A quantum register of arity \textit{n} is a normalized vector in 
$\otimes^{n}_{i=1}\CC^{2}$. We fix an orthonormal basis of $\otimes^{n}_{i=1}\CC^{2}$,
namely $\{ |i> | i \mbox{ is a binary string of length } n \}$.
For example  $1/\sqrt{2} |01> + 1/\sqrt{2} |00>\in \CC^{2}\otimes\CC^{2}$  
is a quantum register of two qubits.

An important property of quantum registers of $n$ qubits is the fact
that it is not always possible to decompose it into 
$n$ isolated qubits (mathematically, this means that we are no able to
describe the global state as the tensor product of the single states).
These particular states are called {\itshape entangled} and enjoy 
properties that we can't find in any object of classical physics.
If (the state of) $n$ qubits are entangled, they behave as connected,
independently from the real physical distance.
The strength of quantum computation is essentially based on the 
existence of entangled states.

\subsection{Hilbert Spaces and Quantum Registers}
Even if Hilbert spaces of the shape $\otimes_{i=1}^{n}\CC^2 (\simeq 
\CC^{2^n})$ are commonly used when defining quantum registers, other
spaces will be defined here. As we will see, they allow to handle very
naturally the interaction between variable names in $\lambda$--terms
and superimposed data.

A \emph{quantum variable set (qvs)} is a finite set of quantum
variables (ranged over by variables like $p$, $r$ and $q$). 

\begin{definition}[\textbf{Hilbert Spaces on $\Vr$}]
  Let $\Vr$ a \textit{qvs} (possibly empty) of cardinality $\#\Vr=n$,
  with $\Hs{\Vr}=\{\phi|\  \phi:\{0,1\}^\Vr\rightarrow\CC \}$ we
  will denote the Hilbert Space of dimension $2^{n}$ equipped with:
  \begin{varvarenumerate}
  \item
    An \emph{inner sum} $+ : \Hs{\Vr}\times\Hs{\Vr}\to\Hs{\Vr}$
    defined by $(\phi+\psi)(f)= \phi(f)+\psi(f)$;
  \item
    A \emph {multiplication by a scalar} $\cdot:\CC\times\Hs{\Vr}\to \Hs{\Vr}$\\
    defined by $(c\cdot \phi)(f)= c\cdot(\phi(f))$;
  \item
    An \emph{inner product} $<\ ,\ >: \Hs{\Vr}\times\Hs{\Vr}\to\CC$\\
    defined by $<\phi,\psi>=\sum_{f\in\{0,1\}^\Vr}\phi(f)^*\psi(f)$.
  \end{varvarenumerate}
\end{definition}

The space is equipped with  the \textit{orthonormal
  basis} $\Bs{\Vr}=\{|f>
:f\in\{0,1\}^\Vr\}.$\footnote{$|f>:\{0,1\}^\Vr \to \CC$ is
  defined by: $  |f>(g)= \left\{ \vseq{ 1\ \mbox{if}\ f = g\\
      0\ \mbox{if}\ f \neq g } \right.  $} We call \textit{standard} such a basis.
For example, the standard basis of  the space $\Hs{\{p,q\}}$ is 
$
\{|p\mapsto 0, q\mapsto 0>, 
|p\mapsto 0, q\mapsto 1>,|p\mapsto 1, q\mapsto 0>,|p\mapsto 1, q\mapsto 1>
\}$.

Let $\Vr'\cap \Vr''=\emptyset$. With $\Hs{\Vr'}\otimes\Hs{\Vr''}$ we
denote the tensor product (defined in the usual way) of
$\Hs{\Vr'}$ and $\Hs{\Vr''}$.
If  $\Bs{\Vr'}= \{|f_i>:
i\le 2^{n-1}\}$ and  $\Bs{\Vr''}= \{|g_j>: j\le2^{m-1}\}$ are the orthonormal bases
respectively of $\Hs{\Vr'}$ and $\Hs{\Vr''}$ then 
$\Hs{\Vr'}\otimes\Hs{\Vr''}$ is equipped with the orthonormal basis
$\{ |f_i>\otimes |g_j> : i\le 2^{n-1}, j\le2^{m-1}\}$.
We will abbreviate $|f>\otimes|g>$ with $|f,g>$.

It is easy to show that if $\Vr'\cap \Vr''=\emptyset$ then there is a
standard \textit{isomorphism} 
$$
\Hs{\Vr'}\otimes\Hs{\Vr''} \stackrel{\normalsize {i}_s}{\simeq} \Hs{\Vr'\cup\Vr''}.
$$
In the rest of the paper we will assume
to work up-to such an isomorphism\footnote{
in particular, if 
$\Qr\in \Hs{\Vr}$, $r\not\in\Vr$ and $|r\mapsto c>\in \Hs{\{r\}}$ 
then\\ $\Qr\otimes|r\mapsto c>$ will denote the element 
$i_s(\Qr\otimes|r\mapsto c>)\in\Hs{\Vr\cup\{r\}}$
}.

As for the case of $\CC^{2^n}$,  we need to define the
notion of a \emph{quantum register}.

\begin{definition}[Quantum Register]
  Let $\Vr$ be a qvs, a \textit{quantum register} is a normalized vector in $\Hs{\Vr}$.
\end{definition}

In particular if $\Qr'\in\Hs{\Vr'}$ and $\Qr''\in\Hs{\Vr''}$ are two
quantum registers, with a little abuse of language (authorized by the
previous stated isomorphism) we will say that $\Qr'\otimes\Qr''$ is a
quantum register in $\Hs{\Vr'\cup\Vr''}$.

Quantum computing is essentially based on the application of unitary
operators to quantum registers.
 A linear operator $U: \Hs{\Vr} \to
\Hs{\Vr}$ is called \textit{unitary} if for all $\phi, \psi \in \Hs{\Vr}$, 
$<U(\phi), U(\psi)>=<\phi,\psi>$
The tensor product of unitary operators is defined as follows:
$(U\otimes V)(\phi\otimes \psi) = U(\phi)\otimes U(\psi)$.


Since we are interested in effective computability, we must restrict
the class of admissible unitary transforms.
Following Bernstein and Vazirani~\cite{BerVa97} 
let us define the set $\PCC$ of poly--time computable complex numbers:
\begin{definition}
  A real number $x\in\RR$ is \emph{polynomial-time computable} (in $\PRR$)
  iff there is a deterministic Turing machine which on input $1^n$ computes
  a binary representation of an integer $m\in\ZZ$ such that
  $|m/2^n-x|\leq 1/2^n$. A complex number $z=x+iy$ is \emph{polynomial-time
    computable} (in $\PCC$) iff $x,y\in\PRR$.\par
\end{definition}

Let $U:\CC^{2^{n}}\rightarrow\CC^{2^{n}}$ be an unitary operator.
U is called computable if $U((\PCC)^{2^{n}})\subseteq(\PCC)^{2^{n}}$.
Let $\mathcal{U}$ be the set of all computable operators; it is immediate 
to observe that $\mathcal{U}$ is  effectively enumerable.
In the rest of the paper we assume to work with a fixed 
effective enumeration $(\mathbf{U}_{i})_{i\lt\omega}$ of $\mathcal{U}$.

\begin{definition}
  A quantum register in $\phi\in\Hs{\Vr}$  is \emph{computable} if
  $\phi:\{0,1\}\to \PCC$.
  A unitary operator $U: \Hs{\Vr} \to \Hs{\Vr}$ is called
  ``computable'' if for each computable quantum register $\phi$,
  $U(\phi)$ is computable.
\end{definition}

Let $U:\CC^{2^{n}}\rightarrow\CC^{2^{n}}$ be a computable operator and 
let $\langle q_{0},\ldots,q_{n-1}\rangle$ be a sequences of distinguished 
variables. $U$ and  $\langle q_{0},\ldots,q_{n-1}\rangle$ induce a computable 
operator $U_{\langle q_{0},\ldots,q_{n-1}\rangle}: \HS(\{q_{0},\ldots,q_{n-1}\})\rightarrow\HS(\{q_{0},\ldots,q_{n-1}\})$ 
defined as follows: if $|f>=|q_{j_0}\mapsto b_{j_0},\ldots,q_{j_{n-1}}\mapsto b_{j_{n-1}}>$ is 
an element of the orthonormal basis of $\HS(\{q_{0},\ldots,q_{n-1}\})$, then
$$
U_{\langle q_{0},\ldots,q_{n-1}\rangle}|f>\stackrel{def}{=}U{|b_{j_0},\ldots,b_{j_{n-1}}>}.
$$

Let $\Vr'=\{q_{i_0},\ldots,q_{i_k} \}\subseteq \Vr$. We naturally extend
(by suitable standard isomorphisms) the unitary operator
$U_{<q_{j_{0}},\ldots,q_{j_{k}}>}: \Hs{\Vr'} \to \Hs{\Vr'}$ to the
unitary operator $U_{<< q_{j_{0}},\ldots,q_{j_{k}}>>}: \Hs{\Vr} \to
\Hs{\Vr}$ that acts as the identity on variables not in $\Vr'$ and as
$U_{< q_{j_{0}},\ldots,q_{j_{k}}>}$  on variables in $\Vr'$.


\begin{example}
  Let us consider the the standard computable  operator
 $\mathbf{cnot}:\CC^2\otimes\CC^2\rightarrow{\CC^2\otimes\CC^2}$.
  Intuitively, the {\bfseries cnot} operator complements the target
  bit (the second one) if the control bit is 1, otherwise does not
  perform any action:

\begin{tabular}{cc}
\begin{minipage}{3cm}
\begin{eqnarray*}
  \mathbf{cnot}|0 0>&=&|0 0>\\
  \mathbf{cnot}|0 1>&=&|0 1>\\
\end{eqnarray*}
\end{minipage}
  &
\begin{minipage}{3cm}
\begin{eqnarray*}
  \mathbf{cnot}|1 0>&=&|1 1>\\
  \mathbf{cnot}|1 1>&=&|1 0>\\
\end{eqnarray*}
\end{minipage}
\end{tabular}

Let us fix the sequence $\langle p, q \rangle$ of variables, 
$\mathbf{cnot}$ induces the operator 
$\mathbf{cnot}_{\langle\langle
  p,q \rangle\rangle} : \HS(\{p,q\}) \rightarrow \HS(\{p,q\})$ such
that:
\begin{eqnarray*}
\mathbf{cnot}_{\langle\langle p,q\rangle\rangle} |q\mapsto 0,p\mapsto
0>&=& |q\mapsto 0, p\mapsto 0>;\\
\mathbf{cnot}_{\langle\langle p,q\rangle\rangle} |q\mapsto 0,
p\mapsto 1 >&=& | q\mapsto 1, p\mapsto 1>;\\
\mathbf{cnot}_{\langle\langle p,q\rangle\rangle} |q\mapsto 1,p\mapsto
0>&=&|q\mapsto 1, p\mapsto 0>;\\
\mathbf{cnot}_{<< p,q>>} |q\mapsto 1,
p\mapsto 1 >&=&| q\mapsto 0, p\mapsto 1>.
\end{eqnarray*}
Please note that 
$|q\mapsto c_1, p\mapsto c_2> = |p\mapsto c_2, q\mapsto c_1> $ (consequently  $\mathbf{cnot}_{<<p,q>>} |q\mapsto c_1, p\mapsto c_2>=
\mathbf{cnot}_{<< p,q>>} |p\mapsto c_2,q\mapsto c_1>$). 
On the other hand, the operators $\mathbf{cnot}_{<< p,q>>}$ and
$\mathbf{cnot}_{<< q,p>>}$ are different: both act as
controlled not, but $\mathbf{cnot}_{<< p,q>>}$ uses $p$ as control bit
while $\mathbf{cnot}_{<< q,p>>}$ uses $q$.

\end{example}

\section{The \qcalc}
Let us associate to each computable unitary operator $\mathbf{U_i}\in {\mathcal U}$ a symbol $U_i$


\subsection*{Terms} 
The set of the \emph{term expressions}, or
\emph{terms} for short, is defined by the following grammar:

$$\begin{array}{lcl}
  x & ::= & v_0, v_1,\ldots  \hfill \mbox{classical variables}\\ 
  r & ::= & r_0,r_1,\ldots   \hfill\mbox{quantum variables}\\ 
  \pi & ::= & x\ \mid\ <x_1,\ldots,x_n>  \hfill \mbox{patterns}\\ 
  B & ::= & 0\ \mid\ 1\   \hfill \mbox{boolean constants}\\ 
  U & ::= & U_0, U_1,\ldots   \hfill \mbox{unitary operators}\\ 
  C & ::= & B\ \mid\ U  \hfill\mbox{constants}\\ 
  M & ::= & x\ \mid\ r\ \mid !(M)\ \mid C \mid \new{M}\ \mid (M_1)M_2\ \mid \\ 
  & & <M_1,\ldots,M_n>\mid\ \lambda!x.M\ \mid\ \lambda \pi.M  \\ 
  & & 
 \hfill \mbox{terms (\mbox{where }  $n\geq 2$)}
\end{array}$$
We assume to work modulo variable renaming, i.e., terms are
equivalence classes modulo $\alpha$-conversion. Substitution up to
$\alpha$-equivalence is defined in the usual way.
Let us denote with $\Qvt{M_1,\ldots,M_k}$ the set of quantum variables
occurring in $M_1,\ldots,M_k$.
Notice that:
\begin{varitemize}
\item
  Variables are either \emph{classical} or \emph{quantum}: the first ones are the 
  usual variables of lambda calculus, while each quantum variable refers to a qubit
  in the underlying quantum register (to be defined shortly).
\item
  There are two sorts of constants as well, namely \emph{boolean constants} ($0$ and $1$)
  and \emph{unitary operators}: the first ones are useful for generating qubits
  and play no role in classical computations, while unitary operators are applied
  to (tuples of) quantum variables when performing quantum computation.
\item
  The term constructor $\new{\cdot}$ creates a new qubit when applied to a 
  boolean constant.
\end{varitemize}
The rest of the calculus is a standard linear lambda calculus, similar
to the one introduced in~\cite{Wad94}. Patterns (and, consequently, lambda abstractions)
can only refer to classical variables. 

There is not any measurement operator in the language. We will comment on that in
Section~\ref{sect:measure}.

\subsection{Judgements and Well--Formed Terms} 
An \emph{environment} $\Gamma$ is a (possibly empty) multiset $\Pi\cup\Delta\cup \Theta$
where $\Pi$ is a (possibly empty) multiset $\pi_1,\ldots,\pi_n$ of
patterns and $\Delta$ is a (possibly empty) multiset
$!x_1,\ldots,!x_n$ (where each $x_i$ is a classical variable), and
$\Theta$ is a (possibly empty) multiset of quantum variables.  
We require that each variable
name occurs at most once in $\Gamma$.
With $!\Gamma$ we denote the environment $!x_1,\ldots,!x_n$
whenever $\Gamma$ is $x_1,\ldots,x_n$.

A \emph{judgment} is an expression $\Gamma \vdash M$, where $\Gamma$
is an environment and $M$ is a term.
\begin{figure*}[!htb]
\dlinea
$$
\begin{array}{r@{\quad}c@{\qquad}l} 
  \urule{}{\vdash C}{const} &
  \urule{}{r\vdash r}{qp-var} & 
  \urule{}{x \vdash x}{classic-var}
\end{array}
\begin{array}{r@{\quad}l}
  \urule{\Gamma \vdash M}{\Gamma ,!x\vdash M}{weak} & 
  \urule{\Gamma ,!x,!y\vdash M}{\Gamma ,!z\vdash M\{z/x,z/y\}}{contr}
\end{array}
$$
$$
\begin{array}{r@{\quad}l}
  \urule{!\Gamma\vdash M}{!\Gamma\vdash !M}{prom} & 
  \urule{\Gamma ,x\vdash M}{\Gamma ,!x\vdash M}{der} 
\end{array}
\begin{array}{r@{\quad}l}
  \urule{\Gamma, x_1,\ldots,x_k\vdash M}{\Gamma, <x_1,\ldots,x_k>\vdash M}{L tens} & 
  \urule{\Gamma_1\vdash M_{1} \cdots\Gamma_k\vdash M_{k}}
  {\Gamma_1,\ldots,\Gamma_k \vdash <M_{1},\ldots, M_{k}>}{R tens}
\end{array}
$$
$$
\begin{array}{r@{\quad}r@{\quad}r@{\quad}l} 
  \urule{\Gamma\vdash M}{\Gamma\vdash \new{M}}{\mathtt{new}} &
  \urule{\Gamma ,\pi\vdash M}{\Gamma \vdash \lambda \pi.M}{\limp I} &
  \urule{\Gamma ,!x\vdash M}{\Gamma \vdash \lambda ! x.M}{\to I} &
  \brule{\Gamma_1\vdash M_{1}}{\Gamma_2\vdash M_{2}}{\Gamma_1,\Gamma_2\vdash (M_{1}) M_{2}}{app} \\[4ex]
\end{array}
$$
\dlinea
\caption{Well Forming Rules}\label{fig:wfr}
\end{figure*}
We say that a judgement $\Gamma\vdash
M$ is \textit{well formed} (\textit{notation:} $\wfj{}\Gamma\vdash M$)
if it is derivable by means of the \textit{well forming rules} in
Figure~\ref{fig:wfr}; with $\wfj{d}\Gamma\vdash M$ we denote that $d$ is a
derivation of the well formed judgement $\Gamma\vdash M$.
If $\Gamma\vdash M$ is \textit{well formed} we say also that the
term $M$ is well formed with respect to the environment $\Gamma$.
We say that a term $M$ is \textit{well formed} if the judgment
$\Qvt{M}\vdash M$ is well formed.
\begin{proposition} 
If a term $M$ is well formed then all the
classical variables in it are bounded.
\end{proposition}
\section{Computations}
A \emph{preconfiguration} is a triple $[\Qr,\QV,M]$ where:
\begin{varitemize}
\item $\Qr\in\Hs{\QV}$;
\item $\QV$ is a finite quantum variable set such that 
$\Qvt{M}\subseteq \QV$;
\item $M$ is a term.
\end{varitemize}
Let $\theta:\QV\rightarrow\QV'$ be a function from a set of quantum
variables $\QV$ to another set of quantum variables $\QV'$.
Then we can extend $\theta$ to any term whose quantum
variables are included in $\QV$: $\theta(M)$ will be
identical to $M$, except on quantum variables, which are 
changed according to $\theta$ itself. Observe that
$\Qvt{\theta(M)}\subseteq\QV'$. Similarly, $\theta$ can
be extended to a function from $\Hs{\QV}$ to $\Hs{\QV'}$ in
the obvious way.
\begin{definition}
Two preconfigurations $[\Qr,\QV,M]$ and $[\Qr',\QV',M']$ are equivalent iff
there is a bijection $\theta:\QV\rightarrow\QV'$ such
that $\Qr'=\theta(\Qr)$ and $M'=\theta(M)$.
If a preconfiguration $C$ is equivalent to $C'$, then we 
will write $C\equiv C'$. The relation $\equiv$ is an equivalence relation.
\end{definition}
A \emph{configuration} is an equivalence class of preconfigurations
modulo the relation $\equiv$. Let $\Conf$ be the set of configurations.
\begin{remark}
The way configurations have been defined, namely quotienting
preconfigurations over $\equiv$, is very reminiscent of usual
$\alpha$-conversion in lambda-terms.
\end{remark}
Let $\redrul=\{\Uq,\nw,\lbeta,\qbeta,\cbeta,\lcom,\rcom,\tupla{i}\}$.
The set $\redrul$ will be ranged over by $\alpha,\beta,\gamma$.
For each $\alpha\in\redrul$, we can define a reduction
relation $\predto{\alpha}\subseteq\Conf\times\Conf$
by means of the rules in Figure~\ref{fig:reduction}. 
\begin{figure*}[!htb]
\dlinea
$$
\begin{array}{l}%
 \mbox{}\hspace{-2ex} 
 \urule
  {
    \vseq{
      {[\Qr,\QV,M_i]\predto{\alpha}[\Qr',\QV',M'_i]}
      \\
      \Ok{[\Qr,\QV,<M_1,\ldots,M_i,\ldots,M_k>]}
    }
  }
  {
    [\Qr,\QV,,<M_1,\ldots,M_i,\ldots,M_k>]\predto{\alpha}
    {[\Qr',\QV',<M_1,\ldots,M'_i,\ldots,M_k>]}
  }%
  {\tupla{i}}
  %
\ \ \hfill 
  \urule
  {
    \vseq{
      [\Qr,\QV,N]\predto{\alpha}[\Qr',\QV',N']
      \\
      \Ok{[\Qr,\QV,MN]} 
    }
  } 
  {[\Qr,\QV,MN]\predto{\alpha}[\Qr',\QV',MN']}
  {\rapp}
  \\[6ex]
  \urule{
    [\Qr,\QV,M]\predto{\alpha}[\Qr',\QV',M']
    \quad \Ok{[\Qr,\QV,MN]}
  } 
  {[\Qr,\QV,MN]\predto{\alpha}[\Qr',\QV',M'N]}
  {\lapp}
  \hfill
  \urule{[\Qr,\QV,M]\predto{\alpha}[\Qr',\QV',M']}
  {[\Qr,\QV,(\lambda \pi . M)]\predto{\alpha}[\Qr',\QV',(\lambda\pi.M')]}
  {\inlambda} \\[4ex]
  \urule{
    \Ok{[\Qr,\QV,U <r_{i_1},...,r_{i_n}>]} 
    \quad \mathbf{U}:\CC^{2^{n}}\rightarrow\CC^{2^{n}}
  }
  {
    \vseq{
      [\Qr,\QV,U<r_{i_1},...,r_{i_n}>]\predto{\Uq}
      \\
      \mbox{}\hspace{10ex}
      {[\mathbf{U}_{<< r_{i_1},\ldots,r_{i_n}>>}Q,\QV,<r_{i_1},...,r_{i_n}>]}
    } 
  }
  {\Uq} 
  %
\hfill
  \urule{
    {\Ok{[\Qr,\QV,\Ln,\new{c}]}\quad r\mbox{ is fresh} }
  } 
  {
    \vseq{
      [\Qr,\QV,\new{c}]\predto{\nw}
      \\
      \mbox{}\hspace{10ex}
      [\Qr\otimes\dr{r\mapsto c},\QV\cup\{r\}, p ]
      }
  } 
  {\nw} \\[6ex]
  \urule{
    \Ok{[\Qr,\QV,(\lambda x.M)N]} 
  } 
  { 
    \vseq{
    [\Qr,\QV,(\lambda x.M)N]\predto{\lbeta}
    \\
      \mbox{}\hspace{10ex}
    [\Qr,\QV,M\{N/x\}]
    }
  }
  {\lbeta} 
  \hfill
  \urule{
    \Ok{[\Qr,\QV,(\lambda<x_1,\ldots,x_n>.M)<r_1,\ldots,r_n>]}
  }
  {
    \vseq{
      [\Qr,\QV,(\lambda<x_1,\ldots,x_n>.M)<r_1,\ldots,r_n>]\predto{\qbeta}
      \\
      \mbox{}\hspace{10ex}
      [\Qr,\QV,M\{r_1/x_1,\ldots,r_n/x_n\}]
    }
  } 
  {\qbeta}
  \\[4ex]
  \mbox{}
\qquad\qquad\qquad\qquad\qquad\qquad
   \urule{ \Ok{[\Qr,\QV,(\lambda! x.M)!N]} } { [\Qr,\QV,(\lambda!
      x.M)!N]\predto{\cbeta}[\Qr,\QV, M\{N/x\}] } {\cbeta}
  \\[4ex]
  \urule {
    \Ok{[\Qr,\QV,L((\lambda \pi.M)N)]}
  }
  {
    \vseq{
      [\Qr,\QV,L((\lambda \pi.M)N)]\predto {\lcom}
      \\
      \mbox{}\hspace{10ex}
      [\Qr,\QV,(\lambda\pi.LM)N]
    }
  } 
  {\lcom}
  %
  \hfill
  \urule {
    \Ok{[\Qr,\QV,((\lambda \pi.M)N)L]}
  }
  {
    \vseq{
      [\Qr,\QV,((\lambda \pi.M)N)L]\predto{\rcom}
      \\
      \mbox{}\hspace{10ex}
      [\Qr,\QV,(\lambda\pi.ML)N]
    }
  } 
  {\rcom}
\end{array}
$$
\dlinea
\caption{Reduction rules.}
\label{fig:reduction}
\end{figure*}
For any subset $\mathscr{S}$ of $\redrul$, we can construct
a relation $\predto{\mathscr{S}}$ by just taking the union
over $\alpha\in\mathscr{S}$ of $\predto{\alpha}$. In particular,
$\redto$ will denote $\predto{\redrul}$. The usual notation
for the transitive and reflexive closures will be used. In particular,
$\rredto$ will denote the transitive and reflexive closure of $\redto$.

Notice we have defined $\redto$ by closing reduction rules under any
context except the ones in the form $!M$. So $\redto$ is not a strategy but,
nevertheless, confluence holds. This is in contrast with $\lambda_{sv}$, where a 
strategy is indeed necessary (even if we do not take into account the
nondeterministic effects of the measurement operator).
\subsection{Subject Reduction}
In this section we propose a Subject Reduction theorem and some related results.
Notice that the calculus is type--free, so Subject Reduction is given
with respect to Well Forming Rules.
\condinc{
\begin{lemma}[Substitution Lemma (linear case)]\label{lemma:linsub}
  For each derivation $d_{1},d_{2}$, if $\wfj{d_{1}}
  \Gamma_{1},x\vdash M$ and $\wfj{d_{2}} \Gamma_{2}\vdash N$, then
  $\wfj{}\Gamma_{1},\Gamma_{2}\vdash M[N/x]$.
\end{lemma}
\begin{proof}
  The proof is by induction on the height of $d_1$ and by cases on the last rule.
  Let $\mathsf{r}$ be the last rule of $d_1$.
  \begin{enumerate}
  \item $\mathsf{r}$ is either \textit{const}, or \textit{qp--var}, or
    \textit{classical--var}: trivial;
  \item $\mathsf{r}$ is $\urule{\Gamma_{1},x\vdash
      M}{\Gamma_{1},!y,x\vdash M}{weak}$.\\
 By IH we have:
    $\wfj{}\Gamma_{1},\Gamma_{2}\vdash M[N/x]$,   
    and by means of  \textit{weak},
   $\wfj{}\Gamma_{1},\Gamma_{2},!y\vdash M[N/x]$
  \item $\mathsf{r}$ is $\urule{\Gamma_{1},x,y\vdash M}{\Gamma_{1},x,!y\vdash M}{der}$.\\
    By IH we have:
    $\wfj{}\Gamma_{1},\Gamma_{2},y\vdash M[N/x]$, 
   and  by means \textit{der}:
    $\wfj{}\Gamma_{1},\Gamma_{2},!y\vdash M[N/x]$
  \item  $\mathsf{r}$ is 
    $\urule{\Gamma_{1},x,!u,!y\vdash M}{\Gamma_{1},x,!z\vdash M[z/u,z/y]}{contr}$.\\
    By IH we have:
    $\wfj{}\Gamma_{1},\Gamma_{2},!u,!y\vdash M[N/x]$
    and by means of \textit{contr}:
    $\wfj{}\Gamma_{1},\Gamma_{2},!z\vdash M[N/x][z/u,z/y]$ 
  \item $\mathsf{r}$ is 
    $\urule{\Gamma_{1},x,y_{1},\ldots,y_{k}\vdash M}{\Gamma_{1},x,\langle y_{1},\ldots,y_{k}\rangle\vdash M}{Ltens}$.\\
    By IH we have:
    $\wfj{}\Gamma_{1},\Gamma_{2},y_{1},\ldots,y_{k}\vdash M[N/x]$,
    and by means of \textit{Ltens}: 
    $\wfj{}\Gamma_{1},\Gamma_{2},\langle y_{1},\ldots,y_{k}\rangle\vdash M[N/x]$
  \item $\mathsf{r}$ is 
      $\brule{\Gamma_{1},x\vdash M_{1}}{\Gamma_{2}\vdash M_{2}}{\Gamma_{11},\Gamma_{12},x\vdash M_{1}M_{2}}{app}$.\\
      By IH we have: $\wfj{}\Gamma_{1},\Gamma_{2}\vdash M_{1}[N/x]$, 
      and by means of \textit{app}:
      $\wfj{}\Gamma_{1},\Gamma_{2}\vdash M_{1}[N/x]$
    \item $\mathsf{r}$ is $\brule{\Gamma_{1}\vdash M_{1}}{\Gamma_{2},
        x\vdash M_{2}}{\Gamma_{11},\Gamma_{12},x\vdash
        M_{1}M_{2}}{app}$.  As for the previous case.
    \item  $\mathsf{r}$ is $\urule{\Gamma_{1},x,!y\vdash M}{\Gamma_{1},x\vdash \lambda !y.M}{\to I}$.\\
    By IH we have:
    $\wfj{}\Gamma_{1},\Gamma_{2},!y\vdash M[N/x]$,
    and by means of $\to I$:
    $\wfj{}\Gamma_{1},\Gamma_{2}\vdash \lambda !y.M[N/x]$
  \item  $\mathsf{r}$ is 
    $\urule{\Gamma_{1},x,\pi\vdash M}{\Gamma_{1},x\vdash \lambda \pi.M}{\limp I} $. As for the  previous case.
  \item  $\mathsf{r}$ is 
    $\urule{\Gamma_{1},x\vdash M}{\Gamma_{1},x\vdash \new{M}}{new}$.\\
    By IH we have: 
    $\wfj{}\Gamma_{1},\Gamma_{2}\vdash M[N/x]$
    and by means of \textit{new}:
    $\wfj{}\Gamma_{1},\Gamma_{2}\vdash \new{M[N/x]}$
  \item $\mathsf{r}$ is $\urule{\Gamma_{11}\vdash
      M_{1},\ldots,\Gamma_{1i},x \vdash M_{i},\ldots,\Gamma_{1k}\vdash
      M_{k}}{\Gamma_{11},\ldots,\Gamma_{1k},x\vdash \langle
      M_{1},\ldots,M_{k}\rangle}{RTens}$.\\  
    By IH we have:
    $\wfj{}\Gamma_{1i},\Gamma_{2}\vdash M_{i}[N/x]$, and by means of
    \textit{RTens}:
    $\wfj{}\Gamma_{11},\ldots,\Gamma_{1k},\Gamma_{2}\vdash\langle
    M_{1},\ldots, M_{i}[N/x],\ldots,M_{k}\rangle$\\ (note that$\langle M_{1},\ldots,
    M_{i}[N/x],\ldots,M_{k}\rangle\equiv \langle
    M_{1},\ldots,M_{k}\rangle [N/x])$
  \end{enumerate}
\end{proof}

\begin{lemma}[Substitution (non linear case)]
  For each derivation $d_{1}, d_{2}$ and for every sequence
  (eventually empty) $x_{1},\ldots,x_{n}$, if $\wfj{d_{1}} \Gamma_{1},
  !x_{1},\ldots,!x_{n}\vdash M$ and $\wfj{d_{2}} !\Gamma_{2}\vdash
  !N$,  then $\wfj{} \Gamma_{1},!\Gamma_{2}\vdash
  M[N/x_{1,},\ldots,N/{x_{n}}]$
\end{lemma}
\begin{proof}
  The proof is by induction on the height of $d_{1}$ and by cases on
  the last rule.  Let $\mathsf{r}$ be the last rule of $d_1$.  We use
  the follow notation: let $\Gamma,!\Delta \vdash !N$ be a judgment,
  we write $\Gamma,!\Delta^{(i)} \vdash !N^{(i)}$ to denote $i$--th
  variant, namely the judgment where we have renamed each bang
  variable $u_{j}$ with the fresh name $u_{j}^{i}$.  It is immediate
  to observe that if $\wfj{d}\Gamma,!\Delta \vdash !N$, then for each
  $i$ there exists a derivation $d^i$ such that
  $\wfj{d^i}\Gamma,!\Delta^{(i)} \vdash !N^{(i)}$, and that is, up to
  renaming of bang variables, identical to $d$.
  \begin{enumerate}

\item $\mathsf r$ is either \textit{const}, or \textit{qp-var}, or \textit{classical-var}. Then, $n=0$, and so we obtain the result by application of weakening  rule.\\
In general we can observe that when $n=0$, namely the sequence is empty, the results follow trivially by application of dereliction rule.\\
In the following case we suppose $n\geq 0$.

\item $\mathsf{r}$ is%
  $\urule{\Gamma_{1},!x_{1},\ldots,!x_{n}\vdash
    M}{\Gamma_{1},!x_{1},\ldots,!x_{n},!x_{n+1}\vdash M}{weak}$. \\%
  By IH, we have $\wfj{}\ \Gamma_{1},\Gamma_{2}\vdash
  M[N/x_{1},\ldots,N/x_{n}]$ and by means of $weak$:
  $\Gamma_{1},\Gamma_{2},!x_{n+1}\vdash M[N/x_{1},\ldots,N/x_{n}]$
\item $\mathsf{r}$ is $der$.
  We must distinguish two different cases.\\
  In the first case $\mathsf{r}$ is:
  $\urule{\Gamma_{1},!x_{1},\ldots,!x_{n-1},x_{n}}{\Gamma_{1},!x_{1},\ldots,!x_{n-1},!x_{n}}{der}$.\\
  So, by IH
  $\wfj{}\Gamma_{1},!\Gamma_{2},x_{n}\vdash{M[N/x_{1},\ldots,N/x_{n-1}]}$
  and by substitution lemma in linear case, we obtain\\
  $\wfj{}\Gamma_{1},!\Gamma_{2},x_{n}\vdash{M[N/x_{1},\ldots,N/x_{n-1},N/x_{n}]}$.\\
  In the second case, we apply dereliction rule on a variable in $\Gamma_{1}$.\\
  Let $\Gamma_{1} = \Gamma_{1}', y$; $\mathsf{r}$ is
  $\urule{\Gamma_{1}',y,!x_{1},\ldots,!x_{n}\vdash
    M}{\Gamma_{1}',!y,!x_{1},\ldots,!x_{n}\vdash M}{der}$.\\
  So, by IH $\wfj{}\Gamma_{1}',y,!\Gamma_{2}\vdash
  M[{N/x_{1},\ldots,N/x_{n}}]$.  Then, by means of $der$
  $\wfj{}\Gamma_{1}',!y,!\Gamma_{2}\vdash M[{N/x_{1},\ldots,N/x_{n}}]$
\item $\mathsf{r}$ is $contr$; as in the previous case, we distinguish
  two case.\\
  If we contract two variables in a variable $!x_{i}$ not in sequence
  $!x_{1},\ldots,!x_{n}$ , we have
  $\urule{\Gamma_{1},!x_{1},\ldots,!x_{n},!y,!z\vdash
    M}{\Gamma_{1},!x_{1},\ldots,!x_{n},!u\vdash M}{contr}$. \\
  By IH we have
  $\wfj{}\Gamma_{1},!\Gamma_{2},!y,!z\vdash{M[N/x_{1},\ldots,N/x_{n}]}$
  and applying the contraction rule on $!y$ and $!z$ we obtain
  $\wfj{}\Gamma_{1},!\Gamma_{2},!u\vdash{M[N/x_{1},\ldots,N/x_{n}][u/y,u/z]}$.
  \\
  Otherwise, if we contract two variables in a variable of sequence,
  we have $\urule{\Gamma_{1},!x_{1},\ldots,!x_{n-1},!y,!z\vdash
    M}{\Gamma_{1},!x_{1},\ldots,!x_{n-1},!x_{n}\vdash M}{contr}$.\\
  By IH. we have 
  $\wfj{}\Gamma_{1},!\Gamma_{2}\vdash
  M[N/x_{1},\ldots,N/x_{n-1},N/y,N/z]$ and the thesis follows
  observing that \\
  $M[N/x_{1},\ldots,N/x_{n-1},N/y,N/z]=M[N/x_{1},\ldots,N/x_{n-1},x_{n}/y,x_{n}/z][N/x_{n}]$.
\item $\mathsf{r}$ is
  $\urule{\Gamma_{1},y_{1},\ldots,y_{k},!x_{1},\ldots,!x_{n}\vdash
    M}{\Gamma_{1},\langle
    y_{1},\ldots,y_{k}\rangle,!x_{1},\ldots,!x_{n}\vdash M}{Ltens}$.\\
  By IH we have
  $\wfj{}\Gamma_{1},y_{1},\ldots,y_{k},!\Gamma_{2}\vdash
  M[N/x_{1},\ldots,N/x_{n}]$ and by means of $Ltens$ we obtain\\
  $\wfj{}\Gamma_{1},\langle
  y_{1},\ldots,y_{k}\rangle,!\Gamma_{2}\vdash
  M[N/x_{1},\ldots,N/x_{n}]$
\item $\mathsf{r}$ is $\urule{ \Gamma_{11},!x_{1},\ldots,!x_{k}\vdash
    M_{1}\ \ \Gamma_{12},!x_{k+1},\ldots,!x_{n}\vdash M_{2} } {
    \Gamma_{11},\Gamma_{12},!x_{1},\ldots,!x_{k},!x_{k+1},\ldots,!x_{n}\vdash
    M_{1}M_{2}}{app}$ and $\Gamma_{1}=\Gamma_{11},\Gamma_{12}$.
  \\
  We use IH with $!\Gamma_{2}^{(1)}\vdash !N^{(1)}$ and
  $!\Gamma_{2}^{(2)}\vdash !N^{(2)}$ as variants of the statement
  $!\Gamma_{2}\vdash !N$ and we obtain \\
  $\wfj{}\Gamma_{11},!\Gamma_{2}^{(1)}\vdash
  M_{1}[N^{(1)}/x_{1},\ldots,N^{(1)}/x_{k}]$ and
  $\wfj{}\Gamma_{12},!\Gamma_{2}^{(2)}\vdash M_{2}[N^{(2)}/x_{k+1},\ldots,N^{(2)}/x_{n}]$.\\
  So, by means of $app$ we have
  \\
  $\wfj{}\Gamma_{11},\Gamma_{12},!\Gamma_{2}^{(1)},!\Gamma_{2}^{(2)}\vdash
  M_{1}[N^{(1)}/x_{1},\ldots,N^{(1)}/x_{k}]M_{2}[N^{(2)}/x_{k+1},\ldots,N^{(2)}/x_{n}]$
  and by several contractions we have thesis.
\item $\mathsf{r}$ is $\urule{\Gamma_{1},!x_{1},\ldots,!x_{n},!y\vdash
    M}{\Gamma_{1},!x_{1},\ldots,!x_{n}\vdash \lambda !y.M}{\to I}$.\\
  By IH $\wfj{}\Gamma_{1},!\Gamma_{2},!y\vdash
  M[N/x_{1},\ldots,N/x_{n}]$ and by means $\to I$ we obtain
  $\wfj{}\Gamma_{1},!\Gamma_{2},\vdash \lambda
  !y. M[N/x_{1},\ldots,N/x_{n}]$
\item $\mathsf{r}$ is $\urule{\Gamma_{1},!x_{1},\ldots,!x_{n},!y\vdash
    M}{\Gamma_{1},!x_{1},\ldots,!x_{n}\vdash \lambda !y.M}{\limp}$.
  As for the previous case.
\item $\mathsf{r}$ is $\urule{\Gamma_{1},!x_{1},\ldots,!x_{n}\vdash
    M}{\Gamma_{1},!x_{1},\ldots,!x_{n}\vdash \new{M}}{new}$.\\
  By IH  $\wfj{}\Gamma_{1},!\Gamma_{2}\vdash M[N/x_{1},\ldots,N/x_{n}]$
  and by means of $new$ we obtain $\wfj{}\Gamma_{1},!\Gamma_{2}\vdash
  \new{M[N/x_{1},\ldots,N/x_{n}]}$
\item $\mathsf{r}$ is
  $\urule{\Gamma_{11},!x_{1},\ldots ,!x_{r}\vdash M_{1}\ldots \Gamma_{1k},!x_{s},\ldots,!x_{n}\vdash M_{k}}{\Gamma_{11},\ldots \Gamma_{1 k},!x_{1}, \ldots,!x_{n}\vdash \langle M_{1},\ldots,M_{k}\rangle}{Rtens}$.\\
  We use IH with $!\Gamma_{2}^{(1)}\vdash !N^{(1)}, \ldots,
  !\Gamma_{2}^{(k)}\vdash !N^{(k)}$
  as  variants of the statement $!\Gamma_{2}\vdash !N$ and we obtain \\
  $\wfj{}\Gamma_{11},\Gamma_{2}^{(1)}\vdash M_{1}[N^{(1)}/x_{1},\ldots,N^{(1)}/x_{r}]$\\
  $\vdots$\\
  $\wfj{}\Gamma_{1k},\Gamma_{2}^{(k)}\vdash
  M_{1}[N^{(k)}/x_{s},\ldots,N^{(k)}/x_{n}]$
  \\
  So, by means of the  tensor rule we have
  $\wfj{}\Gamma_{11},\ldots,\Gamma_{1k},\Gamma_{2}^{(1)},\ldots,\Gamma_{2}^{(k)}\vdash$\\
  $ \langle
  M_{1}[N^{(1)}/x_{1},\ldots,N^{(1)}/x_{r}],\ldots,M_{k}[N^{(k)}/x_{s},\ldots,N^{(k)}/x_{n}]\rangle$,
  and by several application of contractions we obtain thesis.
\item $\mathsf{r}$ is $prom$.  
 In order to apply the  promotion rule,
  $\Gamma_{1}$ must to be $!\Delta$. Therefore $\textsf{r}$ is
  $\urule{!\Delta,!x_{1},\ldots,!x_{n}\vdash
    M}{!\Delta,!x_{1},\ldots,!x_{n}\vdash !M}{prom}$.\\ By IH we have 
  $\wfj{}!\Delta,!\Gamma_{2}\vdash M[N/x_{1},\ldots,N/x_{n}] $ and by
  means of $prom$ $\wfj{}!\Delta,!\Gamma_{2}\vdash
  !M[N/x_{1},\ldots,N/x_{n}] $
  \end{enumerate}
\end{proof}

\begin{lemma}[Substitution (quantum case)] 
 For each derivation
   $d_{1},d_{2}$, for every non empty sequence $x_{1},\dots,x_{n}$, and
   for every non empty sequence $r_{1},\ldots,r_{n}$,\\
   if 
   $\wfj{d_{1}}\Gamma_{1}, \langle x_{1},\ldots,x_{n}\rangle\vdash
  M$ 
  and 
  $\wfj{d_{2}} !\Gamma_{2},r_{1},\ldots,r_{n}\vdash \langle
  r_{1},\ldots,r_{n}\rangle$,\\
   with $r_{1},\ldots,r_{n}\notin M$, then
   $\wfj{}\Gamma_{1},!\Gamma_{2},r_{1},\ldots,r_{n}\vdash
   M[r_{1}/x_{1},\ldots,r_{n}/x_{n}]$
\end{lemma}

\begin{proof} 
The proof is by induction on the height of  $d_{1}$, and by cases on the last rule. Let $\mathsf r$ be the last rule of $d_{1}$.\\
\begin{enumerate}
\item $\mathsf{r}$ is either \textit{const}, or \textit{qp-var}, or
  \textit{classical-var}: trivial;
\item $\mathsf{r}$ is
  $\urule{\Gamma_{1},\langle x_{1},\ldots,x_{n}\rangle \vdash M}{\Gamma_{1},!y,\langle x_{1},\ldots,x_{n}\rangle \vdash M} {weak}$. \\
  By IH we have:
  $\wfj{}\Gamma_{1},!\Gamma_{2},r_{1},\ldots,r_{n}\vdash
  M[r_{1}/x_{1},\ldots,r_{n}/x_{n}]$ and by means of \textit{weak},\\
  $\wfj{}\Gamma_{1},!y,!\Gamma_{2},r_{1},\ldots,r_{n}\vdash
  M[r_{1}/x_{1},\ldots,r_{n}/x_{n}]$
\item $\mathsf{r}$ is $\urule{\Gamma_{1},!u,!y,\langle
    x_{1},\ldots,x_{n}\rangle\vdash M}{\Gamma_{1},!z,\langle
    x_{1},\ldots,x_{n}\rangle\vdash M}{contr}$. \\
  By IH we have: $\wfj{}\Gamma_{1},!u,
  !y,!\Gamma_{2},r_{1},\ldots,r_{n}\vdash
  M[r_{1}/x_{1},\ldots,r_{n}/x_{n}]$, and by means of \textit{contr}:\\
  $\wfj{}\Gamma_{1},!z,!\Gamma_{2},r_{1},\ldots,r_{n}\vdash
  M[r_{1}/x_{1},\ldots,r_{n}/x_{n}]$
\item $\mathsf{r}$ is $\urule{\Gamma_{1},x_{1},\ldots,x_{n}\vdash
    M}{\Gamma_{1},\langle x_{1},\ldots,x_{n}\rangle\vdash M}{Ltens}$.\\
  By means of lemma\ref{lemma:linsub}, applied to
  $\wfj{}\Gamma_{1},x_{1},\ldots,x_{n}\vdash M$ and $r_{1}\vdash
  r_{1}$ we obtain $\Gamma_{1},r_{1},x_{2},\ldots,x_{n}\vdash
  M[r_{1}/x_{1}]$, and by successive applications of  the
  lemma~\ref{lemma:linsub} with respect axioms $r_2\vdash r_2, \ldots,
  r_n\vdash r_n$ we obtain\\ $\wfj{}\Gamma_{1},r_{1},\ldots,r_{n} \vdash
  M[r_{1}/x_{1},\ldots,r_{n}/x_{n}]$.\\
  Then, by several application of weakening, we obtain
  $\wfj{}\Gamma_{1},!\Gamma_{2},r_{1},\ldots,r_{n}\vdash
  M[r_{1}/x_{1},\ldots,r_{n}/x_{n}]$
\item
  Let us suppose that the pattern $\langle x_{1},\ldots,x_{n}\rangle$ belong to the left judgment of the rule $\mathsf{r}$ (the symmetric case is handled in a similar way).\\
  $\mathsf{r}$ is $\brule{\Gamma_{11},\langle
    x_{1},\ldots,x_{n}\rangle\vdash M_{1}} {\Gamma_{12}\vdash
    M_{2}}{\Gamma_{11},\Gamma_{12},\langle x_{1},
    \dots,x_{n\rangle}\vdash M_{1}M_{2}}{app}$.\\
  By IH we have
  $\wfj{}\Gamma_{11},!\Gamma_{2},r_{1},\ldots,r_{n}\vdash
  M_{1}[r_{1}/x_{1},\ldots,r_{n}/x_{n}]$ and by means of \textit{app}:\\
  $\Gamma_{11},\Gamma_{12},!\Gamma_{2},r_{1},\ldots,r_{n}
  \vdash M_{1}[r_{1}/x_{1},\ldots,r_{n}/x_{n}]M_{2}$\\
  Observe that $M_{1}[r_{1}/x_{1},\ldots,r_{n}/x_{n}]M_{2}\equiv
  (M_{1}M_{2})[r_{1}/x_{1},\ldots,r_{n}/x_{n}]$ and conclude.
\item $\mathsf{r}$ is $\urule{\Gamma_{1},\langle
    x_{1},\ldots,x_{n}\rangle,!y\vdash M}{\Gamma_{1},\langle
    x_{1},\ldots,x_{n}\rangle\vdash\lambda !y. M}{\to I}$.\\
  By IH we have\\
  $\wfj{}\Gamma_{1},!y,!\Gamma_{2},r_{1},\ldots,r_{n}\vdash
  M[r_{1}/x_{1},\ldots,r_{n}/x_{n}]$
  and by means of $\to I$:\\
  $\Gamma_{1},!\Gamma_{2},r_{1},\ldots,r_{n}\vdash \lambda
  !y.M[r_{1}/x_{1},\ldots,r_{n}/x_{n}]$
\item $\mathsf{r}$ is $\urule{\Gamma_{1},\langle
    x_{1},\ldots,x_{n}\rangle,\pi\vdash M}{\Gamma_{1},\langle
    x_{1},\ldots,x_{n}\rangle\vdash\lambda \pi. M}{\limp I}$.  As for
  the previous case.
\item $\mathsf{r}$ is $\urule{\Gamma_{1},\langle
    x_{1},\ldots,x_{n}\rangle\vdash M}
  {\Gamma_{1},\langle x_{1},\ldots,x_{n}\rangle\vdash \new{M}}{new}$.\\
  By IH we have $\wfj{}\Gamma_{1},!\Gamma_{2},r_{1},\ldots,r_{n}\vdash
  M[r_{1}/x_{1},\ldots,r_{n}/x_{n}]$ and by means of $\nw$ rule we
  obtain\\
  $\wfj{}\Gamma_{1},!\Gamma_{2},r_{1},\ldots,r_{n}\vdash
  \new{M[r_{1}/x_{1},\ldots,r_{n}/x_{n}]}$
\item $\mathsf{r}$ is
  $\urule{\Gamma_{11}\vdash M_{1}\ \ldots\ 
    \Gamma_{1i},\langle x_{1},\ldots,x_{n}\rangle\vdash 
    M_{i}\ \ldots\ \Gamma_{1k}\vdash M_{k}}
  {\Gamma_{11},\ldots,\Gamma_{1k},\langle x_{1},\ldots,x_{n}\rangle 
    \vdash \langle M_{1},\ldots,M_{k}\rangle}{Rtens}$.\\
  By IH (w.r.t.  $\wfj{}\Gamma_{1i},<x_{1},\ldots,x_{n}>\vdash 
    M_{i}$) we have \\
  $\wfj{}\Gamma_{1i},!\Gamma_{2},r_{1},\ldots,r_{n}\vdash
  M_{1}[r_{1}/x_{1},\ldots,r_{n}/x_{n}]$
  and by means of $R tens$ we conclude\\
  $\wfj{}\Gamma_{11},\ldots,\Gamma_{1k},!\Gamma_{2},r_{1},\ldots r_{n}\vdash\langle M_{1},\ldots,M_{i}[r_{1}/x_{1},\ldots,r_{n}/x_{n}],\ldots,M_{k}\rangle$,\\
  (observe that $\langle
  M_{1},\ldots,M_{i}[r_{1}/x_{1},\ldots,r_{n}/x_{n}],\ldots,M_{k}\rangle\equiv
  \langle M_{1},\ldots,M_{k}\rangle [r_{1}/x_{1},\ldots,r_{n}/x_{n}]$).

\vspace{0.5 cm}
Note that last rule can't be a promotion rule.
\end{enumerate}
\end{proof}
}
{}

\begin{theorem}[Subject Reduction] If  $\wfj{}\Gamma\vdash M$ 
and $[\Qr,\QV, M]\redto[\Qr',\QV', M']$ then $\wfj{}\ \Gamma,
\QV'-\QV\vdash M'$.
\end{theorem}
\condinc{
\begin{proof} 
  The proof is by induction on the height of $d$ and by cases on the
  last rule of $d$.  Let $\mathsf{r}$ be the last rule of $d$.

\begin{enumerate}
\item $\mathsf r$ is either \textit{const}, or \textit{qp-var}, or
  \textit{classical-var}: the proof is trivial.
\item $\mathsf r$ is $app$ and the transition rule is
  \brule{[\Qr,\QV,M_{1}]\predto{\alpha}[\Qr',\QV',M_{1}']}
  {\Ok{[\Qr,\QV,M_{1}M_{2}]}}
  {[\Qr,\QV,M_{1}M_{2}]\predto{\alpha}[\Qr',\QV',M_{1}'M_{2}]}
  {\lapp}\\
  We have $\brule{\Gamma_{1}\vdash M_{1}'}{\Gamma_{2}\vdash
    M_{2}}{\Gamma_{1},\Gamma_{2}\vdash M_{1}'M_{2}}{app}$, so by IH we
  have $\wfj{}\ \Gamma_{1},\QV'-\QV\vdash M_{1}'$, 
and by means of $app$ we obtain 
$\wfj{}\ \Gamma_{1},\Gamma_{2},\QV'-\QV\vdash M_{1}'M_{2}$.
\item  $\mathsf r$ is $app$ and the transition rule 
 \brule
  {[\Qr,\QV,M_{2}]\predto{\alpha}[\Qr',\QV',M_{2}']}
  {\Ok{[\Qr,\QV,M_{1}M_{2}]}} 
  {[\Qr,\QV,M_{1}M_{2}]\predto{\alpha}[\Qr',\QV',M_{1}M_{2}']}
  {\rapp} : simmetric to previous case.
\item $\mathsf r$ is $app$ and the transition rule is $
  \urule{\Ok{[\Qr,\QV,(\lambda x.M)N]} } { [\Qr,\QV,(\lambda
    x.M)N]\predto{\lbeta}[\Qr,\QV,M\{N/x\}]}{\lbeta}$ (application
  generates a redex).  Suppose we have the follow derivation $d$:
$$\brule{\urule{\lto{\Gamma_{1},x\vdash M}{d_{1}}}{\Gamma_{1}\vdash \lambda x.M}{}}{\lto{\Gamma_{2}\vdash N}{d_{2}}}{\Gamma_{1},\Gamma_{2}\vdash (\lambda x.M)N}{}$$
Let  $\wfj{d_{1}}\ \Gamma_{1},x\vdash M$ and $\wfj{d_{2}}\ \Gamma_{2}\vdash N$ .\\
Considering the transition
$[\Qr,\QV,(\lambda x.M)N]\predto{\lbeta}[\Qr,\QV,M[N/x]]$.\\
We note that the transition doesn't modify $\QV$ set, so we've just to
apply substitution lemma on $d_{1}$ and $d_{2}$:
$\wfj{}{\Gamma_{1},\Gamma_{2}\vdash M[N/x]}$.
\item $\mathsf r$ is $app$ and the transition rule is $\qbeta$ or
  $\cbeta$. Similar to previous case.
\item $\mathsf r$ is $app$ and the transition rule is $\urule
  {\Ok{[\Qr,\QV,L((\lambda p.M)N)]}} {[\Qr,\QV,L((\lambda
    \pi.M)N)]\predto{\lcom}[\Qr,\QV,(\lambda\pi.LM)N]} {\lcom}$.\\[2ex]
  Note that the transition rule doesn't modify $\Qr$ and $\QV$.
  So, from derivation:
  $$\brule{\lto{\Gamma_{1}\vdash L}{d_{1}}}{\brule{\urule{\lto{\Gamma_{2}',\pi\vdash M}{d_{2}}}{\Gamma_{2}'\vdash \lambda \pi . M}{\limp I}}{\lto{\Gamma_{2}''\vdash N}{d_{3}}}{\Gamma_{2}\vdash (\lambda \pi . M)N}{app}}{\Gamma_{1},\Gamma_{2}\vdash L((\lambda \pi . M)N)}{app}$$
  we exhibit a derivation of $\Gamma_{1},\Gamma_{2}\vdash
  (\lambda\pi.LM)N$:
   $$\brule{\urule{\brule{\lto{\Gamma_{1}\vdash L}{d_{1}}}{\lto{\Gamma_{2}',\pi\vdash M}{d_{2}}}{\Gamma_{1},\Gamma_{2}',\pi\vdash LM}{app}}{\Gamma_{1},\Gamma_{2}'\vdash\lambda \pi.LM}{\limp I}}{\lto{\Gamma_{2}''\vdash N}{d_{3}}}{\Gamma_{1},\Gamma_{2}\vdash (\lambda\pi.LM)N}{app}$$
 \item $\mathsf r$ is $app$ and the transition rule is $\urule
   {\Ok{[\Qr,\QV,((\lambda p.M)N)L]}} {[\Qr,\QV,((\lambda
     \pi.M)N)L]\predto{\rcom}[\Qr,\QV,(\lambda\pi.ML)N]} {\rcom}$. \\ 
As   in previous case,
$$\brule{\brule{\urule{\lto{\Gamma_{1}',\pi\vdash M}{d_{1}}}{\Gamma_{1}'\vdash \lambda\pi.M}{\limp I}}{\lto{\Gamma_{1}''\vdash N}{d_{2}}}{\Gamma_{1}\vdash (\lambda\pi.M)N}{app}}{\lto{\Gamma_{2}\vdash L}{d_{3}}}{\Gamma_{1},\Gamma_{2}\vdash ((\lambda \pi . M)N)L}{app}$$
then
$$\brule{\urule{\brule{\lto{\Gamma_{1}',\pi\vdash M}{d_{1}}}{\lto{\Gamma_{2} \vdash L}{d_{3}}}{\Gamma_{1}',\Gamma_{2},\pi\vdash ML}{app}}{\Gamma_{1}',\Gamma_{2}\vdash \lambda\pi . ML}{\limp I}}{\lto{\Gamma_{1}''\vdash N}{d_{2}}}{\Gamma_{1},\Gamma_{2}\vdash ((\lambda \pi . ML)N)}{app}$$
\item  $\mathsf r$  is $\limp I$:
$$\urule{\lto{\Gamma,\pi\vdash M}{d_{
      1}}}{\Gamma\vdash \lambda\pi.M}{}$$ If we have
$$\urule{[\Qr,\QV,M]\redto[\Qr',\QV',M']}{[\Qr,\QV,\lambda\pi.M]\redto[\Qr',\QV',\lambda\pi.M']}{}$$
by IH on $d_{1}$ $$\wfj{}\  \Gamma,\pi,\QV'-\QV\vdash M'$$
and we conclude $$\wfj{}\  \Gamma,\pi,\QV'-\QV\vdash \lambda.\pi M'$$
\item  $\mathsf r$ is
$$\urule{{!\Gamma\vdash c}}{!\Gamma\vdash \new{c}}{}$$
We have the following transition rule:
$$[\Qr,\QV,\new{c}]\redto[\Qr\otimes | p\leftarrow c\rangle,\QV\cup\{p\},p]$$
Beginning from axiom 
$$\urule{}{p\vdash p}{}$$
we obtain the result by several application of weakening rule
$$\LT{!\Gamma,p\vdash p}{p\vdash p}{}$$
\item $\mathsf{r}$ is 
$$\urule{{!\Gamma\vdash M}}{!\Gamma\vdash \new{M}}{}$$
in which the argument is a term M.
In this case the proof is in practice identical to the case of application.
\item  $\mathsf r$ is $\urule{\Gamma_1\vdash M_{1} \cdots\Gamma_k\vdash M_{k}}
  {\Gamma_1,\ldots,\Gamma_k \vdash <M_{1},\ldots, M_{k}>}{R tens}$ and  the  transition rule is :

$
\brule
{
  {[\Qr,\QV,M_i]\predto{\alpha}[\Qr',\QV',M'_i]} 
}
{\mbox{}\!\!\!
  \Ok{[\Qr,\QV,<M_1,\ldots,M_i,\ldots,M_k>]}\\
} 
{
  [\Qr,\QV,,<M_1,\ldots,M_i,\ldots,M_k>]\predto{\alpha}[\Qr',\QV',,<M_1,\ldots,M'_i,\ldots,M_k>]
}
{}
$

Then, if we have $\wfj{d_{1}}\ \Gamma_{1},\ldots,\wfj{d_{k}}\ \Gamma_{k}$

by HI, $\wfj{}\ \Gamma_{i},\QV'-QV\vdash M_{i} $ and by means of $Rtens$ $\wfj{}\ \Gamma_{1},\ldots,\QV'-\QV,\ldots,\Gamma_{k}\vdash M_{1},\ldots, M_{k} $ 
\end{enumerate}
\end{proof}
}
{}

\begin{corollary}If  $\wfj{}\Gamma\vdash M$ 
and $[\Qr,\QV, M]\rredto[\Qr',\QV', M']$ then $\wfj{}\ \Gamma,
\QV'-\QV\vdash M$.
\end{corollary}

Notice that $\QV'-\QV$ is  the (possibly empty) set of  new quantum variables
added to $\QV$  by the reduction.

In the following, we will  work with \textit{well--formed} configuration:
\begin{definition}
  A configuration $[\Qr,\QV,M]$ is said to be \textit{well--formed} if there
  is a context $\Gamma$ such that $\Gamma\vdash M$ is well-formed.
\end{definition}
As a consequence of Subject Reduction, the set of well--formed
configurations is closed under reduction:
\begin{corollary} If $M$ is well formed and $[\Qr,\QV, M]
\rredto[\Qr',\QV', M']$ then $M'$ is well formed.
\end{corollary}
In the following, with \textit{configuration} we will
mean \textit{ well--formed configuration}.
Now, let us give the definitions of normal form, configuration and computation.
\begin{definition} A configuration $C =[\Qr,\QV, M]$ is said to be
in \textit{normal form} iff there is no $C'$ such that $C\redto C'$.  Let us
denote with $\NF$ the set of configurations in normal form.
\end{definition}

We define a computation as a suitable sequence of configurations:
\begin{definition} Let $C =[\Qr,\QV, M]$ be a configuration. A
\emph{computation} of length $\varphi\leq\omega$ starting with $C_0$ is a sequence
of configurations $\{C_i\}_{i\lt\varphi}$ such that for all $0\lt i\lt \varphi$,
$C_{i-1}\redto C_{i}$ and either $\varphi=\omega$ or $C_{l-1}\in
\NF$.
\end{definition}


If a computation starts with a
configuration $[\Qr_{0},\QV_{0},M_{0}]$ in which $M_{0}$ does not
contain quantum variables and the set $\QV_0$ is empty, then at each
step $i$ the set $\QV_{i}$ coincides with the set
$\Qvt{M_{i}}$:
\begin{proposition} Let $\{ [\Qr_i,\QV_i, M_i]\}_{i\lt l}$ be a
  computation, such that $\Qvt{M_0} =\emptyset$. Then
  for every $i\lt\varphi$ we have $\QV_i = \Qvt{M_i}$.
\end{proposition}
\condinc{
  \begin{proof}
    Observe that if $[\Qr,\Qvt{M}, M]\redto[\Qr',\QV', M']$ then by
    inspection of reduction rules we immediately have  that $\QV' = \Qvt{M'}$, and
    conclude.
  \end{proof}
}
{}
In the rest of the paper, $[\Qr,M]$ denotes the configuration 
$[\Qr,\Qvt{M},M]$.

\subsection{Confluence}

\condinc{%
  In this section we
  will work with both preconfigurations and configurations (in
  particular with preconfiguration we mean well--formed
  preconfiguration, where the notion of well--formed preconfiguration is
  the same of well--formed configuration).  
  With $\CONF$ we denote the set of well--formed
  preconfigurations.

  The reduction relation $C\predto{\alpha} C'$ between
  preconfigurations is defined as for configurations.  

  If $C, C'$ are configurations (remember that they are equivalence
  classes) such that $C\predto{\alpha} C'$ and  $C^*\in C$ is a
  preconfiguration, then there is a preconfiguration $C^o\in C'$
  s.t. $C^*\predto{\alpha} C^o$. On the other side if $C^*, C^o$ are
  preconfigurations such that $C^*\predto{\alpha} C^o$, then
  $C\predto{\alpha} C'$ where $C,C'$ are the equivalence classes
  respectively of $C^*, C^o$.

  }%
{}

Commutative reduction steps behave very differently to other
reduction steps when considering confluence. As a consequence, it is 
useful to define two subsets of $\redrul$ as follows:
\begin{definition}
We distinguish two particular subsets of $\redrul$, namely
$\commrul=\{\rcom,\lcom\}$ and $\noncommrul=\redrul-\commrul$.
\end{definition}

\condinc{
The following two lemmas refer to preconfigurations.
\begin{lemma}[Uniformity]\label{lemma:uniformity}
  For every $M,M'$ such that $M\predto{\alpha} M'$, exactly one of the
  following conditions holds:
  \begin{varenumerate}
    \item\label{firstcase}
      $\alpha\neq\nw$ and there is a unitary transformation
      $G_{M,M'}:\Hs{\Qvt{M}}\rightarrow\Hs{\Qvt{M}}$ 
      such that $[\Qr,\QV,M]\predto{\alpha}[\Qr',\QV',M']$
      iff $\OK{[\Qr,\QV,M]}$, $\QV'=\QV$ and
      $\Qr'=(G_{M,M'}\otimes I_{\QV-\Qvt{M}})\Qr$.
    \item\label{secondcase}
      $\alpha=\nw$ and there are a constant $c$ and a
      quantum variable $r$ such that 
      $[\Qr,\QV,M]\predto{\nw}[\Qr',\QV',M']$
      iff $\OK{[\Qr,\QV,M]}$, $\QV'=QV\cup\{r\}$ and
      $\Qr'=Qr\otimes\dr{r\leftarrow c}$. Moreover,
      $[\Qr,\QV,M]\predto{\nw}[\Qr\otimes\dr{r'\leftarrow c},
      \QV\cup\{r'\},M'\{r'/r\}]$ whenever $\OK{[\Qr,\QV,M]}$ and 
      $r'\notin\QV$.
  \end{varenumerate}
\end{lemma} 

\begin{proof} 
We go by induction on $M$. $M$ cannot be a variable nor a constant nor
a unitary operator. If $M$ is an abstraction $\lambda\pi.N$, then
$M'=\lambda\pi.N'$, $N\predto{\alpha} N'$ and the thesis follows from the inductive
hypothesis. Similarly when $M=\lambda !x.N$.
If $M=NL$, then we distinguish a number of cases:
\begin{varitemize}
\item
  $M'=N'L$ and $N\predto{\alpha}N'$. The thesis follows from the inductive hypothesis.
\item
  $M'=NL'$ and $L\predto{\alpha}L'$. The thesis follows from the inductive hypothesis.
\item
  $N=U^n$, $L=<r_{i_1},...,r_{i_n}>$ and $M'=<r_{i_1},...,r_{i_n}>$. 
  Then case~\ref{firstcase} holds. In particular,
  $\Qvt{M}=\{r_{i_1},...,r_{i_n}\} $ and $G_{M,M'}=U_{r_{i_1},...,r_{i_n}}$.
\item
  $N=\lambda x.P$ and $M'=P\{L/x\}$. Then case~\ref{firstcase} holds. In particular
  $G_{M,M'}=I_{\Qvt{M}}$.
\item
  $N=\lambda <x_1,\ldots,x_n>.P$, $L=<r_1,\ldots,r_n>$ 
  and $M'=P\{r_1/x_1,\ldots,r_n/x_n\}$. Then case~\ref{firstcase} holds and
  $G_{M,M'}=I_{\Qvt{M}}$.
\item
  $N=\lambda !x.P$, $L=!Q$ 
  and $M'=P\{Q/x\}$. Then case~\ref{firstcase} holds and $G_{M,M'}=I_{\Qvt{M}}$.
\item
  $L=(\lambda \pi.P)Q$ and $M'=(\lambda\pi.NP)Q$. Then case~\ref{firstcase} holds 
  and $G_{M,M'}=I_{\Qvt{M}}$.
\item
  $N=(\lambda \pi.P)Q$ and $M'=(\lambda\pi.PL)Q$. Then case~\ref{firstcase} holds 
  and $G_{M,M'}=I_{\Qvt{M}}$.
\end{varitemize}
If $M=\new{c}$ then $M'$ is a quantum variable $r$ and case~\ref{secondcase} holds.
This concludes the proof.
\end{proof}
}
{}

\condinc{
\begin{lemma}\label{lemma:paramuniformity}
Suppose $[\Qr,\QV,M]\predto{\alpha}[\Qr',\QV',M']$.
\begin{varenumerate}
\item\label{firstclaim}
  If $\OK{[[\Qr,\QV,M\{N/x\}]]}$, then
  $[\Qr,\QV,M\{N/x\}]\predto{\alpha}[\Qr',\QV',M'\{N/x\}]$.
\item\label{secondclaim}
  If $\OK{[[\Qr,\QV,M\{r_1/x_1,\ldots,r_n/x_n\}]]}$,
  then $[\Qr,\QV,M\{r_1/x_1,\ldots,r_n/x_n\}]\predto{\alpha}[\Qr',\QV',M\{r_1/x_1,\ldots,r_n/x_n\}]$
\item\label{thirdclaim}
  If $x,\Gamma\vdash N$ and $\OK{[[\Qr,\QV,N\{M/x\}]]}$,
  then $[\Qr,\QV,N\{M/x\}]\predto{\alpha}[\Qr',\QV',N\{M'/x\}]$
\end{varenumerate}
\end{lemma}
\begin{proof}
Claims~\ref{firstclaim} and~\ref{secondclaim} can be proved by induction on the
proof of  $[\Qr,\QV,M]\predto{\alpha}[\Qr',\QV',M']$. Claim~\ref{thirdclaim} can be
proved by induction on $N$.
\end{proof}
}
{}

Strictly speaking, one-step confluence does not hold in the
\qcalc. For example, 
if $\Ok{[\Qr,\QV,(\lambda\pi.M)((\lambda x.N)L)]}$, then
both
$$
\begin{array}{l}
[\Qr,\QV,(\lambda\pi.M)((\lambda x.N)L)]\predto{\noncommrul}\\
\hspace{0pt}[\Qr,\QV,(\lambda\pi.M)(N\{x/L\})]
\end{array}
$$
and
$$
\begin{array}{l}
[\Qr,\QV,(\lambda\pi.M)((\lambda x.N)L)]\predto{\commrul}\\
\hspace{0pt}[\Qr,\QV,(\lambda x.(\lambda\pi.M)N)L]\predto{\noncommrul}\\
\hspace{0pt}[\Qr,\QV,(\lambda\pi.M)(N\{x/L\})]
\end{array}
$$
However, this phenomenon is only due to the presence of
commutative rules:
\condinc{%
\begin{lemma}[One-step Confluence for preconfigurations]\label{lemma:onestepconf}
  Let $C,D,E$ be preconfigurations with  
  $C\predto{\alpha} D$, $C\predto{\beta}E$ and $D\neq E$.
\begin{varenumerate}
\item
  If $\alpha\in\commrul$ and $\beta\in\commrul$, then there is $F$ 
  with $D\predto{\commrul}F$ and $E\predto{\commrul}F$.
\item 
  If $\alpha\in\noncommrul$ and $\beta\in\noncommrul$, then there is $F$ 
  with $D\predto{\noncommrul}F$ and $E\predto{\noncommrul}F$.
\item 
  If $\alpha\in\commrul$ and $\beta\in\noncommrul$, then either 
  $D\predto{\noncommrul}E$ or there 
  is $F$ with $D\predto{\noncommrul} F$ and $E\predto{\commrul}F$.
\end{varenumerate}
\end{lemma}
\begin{proof}
Let $C=[\Qr,QV,M]$. We go by induction on $M$. $M$ cannot be a variable 
nor a constant nor a unitary operator. If $M$ is an abstraction $\lambda\pi.N$, then
$D=[\Qr',\QV',\lambda \pi.N']$, $D'=[\Qr'',\QV'',\lambda \pi.N'']$ and
\begin{eqnarray*}
  [\Qr,\QV,N]&\predto{\alpha}&[\Qr',\QV',N']\\
  \ [\Qr,\QV,N]&\predto{\beta}&[\Qr'',\QV'',N'']
\end{eqnarray*}
The induction hypothesis easily leads to the thesis. If $M=NL$, 
we can distinguish a number of distinct cases depending on the 
last rule used to prove $C\predto{\alpha}D$, $C\predto{\beta}E$:
\begin{varitemize}
\item
  $D=[\Qr',\QV',N'L]$ and $E=[\Qr'',\QV'',NL']$ where 
  $[\Qr,\QV,N]\predto{\alpha}[\Qr',\QV',N']$
  and $[\Qr,\QV,L]\predto{\beta}[\Qr'',\QV'',L']$. 
  We need to distinguish four sub-cases:
  \begin{varitemize}
  \item
    If $\alpha,\beta=\nw$, then, by Lemma~\ref{lemma:uniformity}, there exist
    two quantum variables $r',r''\notin\QV$ and two constants $c',c''$ such that 
    $\QV'=\QV\cup\{r'\}$, $\QV''=\QV\cup\{r''\}$, 
    $\Qr'=\Qr\otimes\dr{r'\leftarrow c'}$
    and $\Qr''=\Qr\otimes\dr{r''\leftarrow c''}$.
    Applying~\ref{lemma:uniformity} again, we obtain
    \begin{eqnarray*}
      D&\predto{\nw}&[\Qr\otimes\dr{r'\leftarrow c'}\otimes\dr{r'''\leftarrow c''},
      \QV\cup\{r',r'''\},N'L'\{r'''/r''\}]=F\\
      E&\predto{\nw}&[\Qr\otimes\dr{r''\leftarrow c''}\otimes\dr{r''''\leftarrow c'},
      \QV\cup\{r'',r''''\},N'\{r''''/r'\}L']=G
    \end{eqnarray*}
    As can be easily checked, $F\equiv G$.
  \item
    If $\alpha=\nw$ and $\beta\neq\nw$, then, by Lemma~\ref{lemma:uniformity}
    there exists a quantum variable $r$ and a constant $c$ such that
    $\QV'=\QV\cup\{r\}$, $\Qr'=\Qr\otimes\dr{r\leftarrow c}$,
    $\QV''=\QV$ and $\Qr''=(G_{L,L'}\otimes I_{\QV-{\Qvt{L}}})\Qr$. As a consequence,
    applying Lemma~\ref{lemma:uniformity} again, we obtain
    \begin{eqnarray*}
      D&\predto{\beta}&[(G_{L,L'}\otimes I_{\QV\cup\{r\}-{\Qvt{L}}})(\Qr\otimes\dr{r\leftarrow c}),\QV\cup\{r\},N'L']=F\\
      E&\predto{\nw}&[((G_{L,L'}\otimes I_{\QV-{\Qvt{L}}})\Qr)\otimes\dr{r\leftarrow c},\QV\cup\{r\},N'L']=G
    \end{eqnarray*}
    As can be easily checked, $F=G$.
  \item
    If $\alpha\neq\nw$ and $\beta=\nw$, then we can proceed as in the previous
    case.
  \item
    If $\alpha,\beta\neq\nw$, then by Lemma~\ref{lemma:uniformity}, there exist
    $\QV''=\QV'=\QV$, 
    $\Qr'=(G_{N,N'}\otimes I_{\QV-{\Qvt{N}}})\Qr$ and
    $\Qr''=(G_{L,L'}\otimes I_{\QV-{\Qvt{L}}})\Qr$.
    Applying~\ref{lemma:uniformity} again, we obtain
    \begin{eqnarray*}
      D&\predto{\beta}&[(G_{L,L'}\otimes I_{\QV-{\Qvt{L}}})((G_{N,N'}\otimes I_{\QV-{\Qvt{N}}})\Qr),\QV,N'L']=F\\
      E&\predto{\alpha}&[(G_{N,N'}\otimes I_{\QV-{\Qvt{L}}})((G_{L,L'}\otimes I_{\QV-{\Qvt{L}}})\Qr),\QV,N'L']=G
    \end{eqnarray*}
    As can be easily checked, $F=G$.
  \end{varitemize}
\item
  $D=[\Qr',\QV',N'L]$ and $E=[\Qr'',\QV'',N''L]$ where $[\Qr,QV,N]\redto[\Qr',QV',N']$
  and $[\Qr,\QV,N]\redto[\Qr'',\QV'',N'']$. Here we can apply the inductive hypothesis.
\item
  $D=[\Qr',\QV',NL']$ and $E=[\Qr'',\QV'',NL'']$ where $[\Qr,QV,L]\redto[\Qr',QV',L']$
  and $[\Qr,\QV,L]\redto[\Qr'',\QV'',L'']$. Here we can apply the inductive hypothesis
  as well.
\item $N=(\lambda x.P)$, $D=[\Qr,\QV,P\{L/x\}]$, $E=[\Qr',\QV',NL']$,
  where
  $[\Qr,\QV,L]\predto{\beta}[\Qr',\QV',L']$.\\
  Clearly $\Ok{[\Qr,\QV,P\{L/x\}]}$ and, by
  Lemma~\ref{lemma:paramuniformity},
  $[\Qr,\QV,P\{L/x\}]\redto[\Qr',\QV',P\{L'/x\}]$.\\
  Moreover, $[\Qr',\QV',NL']=[\Qr',\QV',(\lambda
  x.P)L']\redto[\Qr',\QV',P\{L'/x\}]$
\item $N=(\lambda x.P)$, $D=[\Qr,\QV,P\{L/x\}]$,
  $E=[\Qr',\QV',(\lambda x.P')L]$, where
  $[\Qr,\QV,P]\predto{\beta}[\Qr',\QV',P']$. Clearly
  $\Ok{[\Qr,\QV,P\{L/x\}]}$ and, by Lemma~\ref{lemma:paramuniformity},
  $[\Qr,\QV,P\{L/x\}]\predto{\beta}[\Qr',\QV',P'\{L/x\}]$.\\
  Moreover, $[\Qr',\QV',(\lambda
  x.P')L]\predto{\beta}[\Qr',\QV',P'\{L/x\}]$
\item $N=(\lambda !x.P)$, $L=!Q$, $D=[\Qr,\QV,P\{Q/x\}]$,
  $E=[\Qr',\QV',(\lambda !x.P')L]$, where
  $[\Qr,\QV,P]\predto{\beta}[\Qr',\QV',P']$. Clearly
  $\Ok{[\Qr,\QV,P\{Q/x\}]}$ and, by Lemma~\ref{lemma:paramuniformity},
  $[\Qr,\QV,P\{Q/x\}]\predto{\beta}[\Qr',\QV',P'\{Q/x\}]$.\\
  Moreover, $[\Qr',\QV',(\lambda
  x.P')!Q]\predto{\beta}[\Qr',\QV',P'\{Q/x\}]$
\item
  $N=(\lambda <x_1,\ldots,x_n>.P)$, $L=<r_1,\ldots,r_n>$, 
  $D=[\Qr,\QV,P\{r_1/x_1,\ldots,r_n/x_n\}]$, $E=[\Qr',\QV',(\lambda <x_1,\ldots,x_n>.P')L]$, where
  $[\Qr,\QV,P]\predto{\beta}[\Qr',\QV',P']$. Clearly 
  $\Ok{[\Qr,\QV,P\{r_1/x_1,\ldots,r_n/x_n\}]}$
  and, by Lemma~\ref{lemma:paramuniformity}, \\
  $[\Qr,\QV,P\{r_1/x_1,\ldots,r_n/x_n\}]\predto{\beta}[\Qr',\QV',P'\{r_1/x_1,\ldots,r_n/x_n\}]$.\\ Moreover, 
  $[\Qr',\QV',(\lambda <x_1,\ldots,x_n>.P')L]\predto{\beta}[\Qr',\QV',P'\{r_1/x_1,\ldots,r_n/x_n\}]$.
\item
  $N=(\lambda x.P)Q$, $D=[\Qr,\QV,(\lambda x.PL)Q]$,
  $E=[\Qr,\QV,(P\{Q/x\})L]$, $\alpha=\rcom$, $\beta=\lbeta$.\\
 Clearly, 
  $[\Qr,\QV,(\lambda x.PL)Q]\predto{\lbeta}[\Qr,\QV,(P\{Q/x\})L]$.
\item
  $N=(\lambda\pi.P)Q$, $D=[\Qr,\QV,(\lambda \pi.PL)Q]$,
  $E=[\Qr',\QV',((\lambda\pi.P')Q)L]$, $\alpha=\rcom$, where 
  $[\Qr,\QV,P]\predto{\beta}[\Qr',\QV',P']$. Clearly, 
  $[\Qr,\QV,(\lambda x.PL)Q]\predto{\rcom}[\Qr',\QV',(\lambda x.P'L)Q]$ and
  $[\Qr',\QV',((\lambda\pi.P')Q)L]\predto{\beta}[\Qr',\QV',(\lambda\pi.P'L)Q]$.
\item
  $N=(\lambda\pi.P)Q$, $D=[\Qr,\QV,(\lambda x.PL)Q]$,
  $E=[\Qr',\QV',((\lambda\pi.P)Q')L]$, $\alpha=\rcom$, where 
  $[\Qr,\QV,Q]\predto{\beta}[\Qr',\QV',Q']$. Clearly, 
  $[\Qr,\QV,(\lambda x.PL)Q]\predto{\rcom}[\Qr',\QV',(\lambda x.PL)Q']$ and
  $[\Qr',\QV',((\lambda\pi.P)Q')L]\predto{\beta}[\Qr',\QV',(\lambda\pi.PL)Q']$.
\item
  $N=(\lambda\pi.P)Q$, $D=[\Qr,\QV,(\lambda x.PL)Q]$,
  $E=[\Qr',\QV',((\lambda\pi.P)Q)L']$, $\alpha=\rcom$, where 
  $[\Qr,\QV,L]\predto{\beta}[\Qr',\QV',L']$. Clearly, 
  $[\Qr,\QV,(\lambda x.PL)Q]\predto{\rcom}[\Qr',\QV',(\lambda x.PL')Q]$ and
  $[\Qr',\QV',((\lambda\pi.P)Q)L']\predto{\beta}[\Qr',\QV',(\lambda\pi.PL')Q]$.
\item
  $N=(\lambda\pi.P)$, $L=(\lambda x.Q)R$, 
  $D=[\Qr,\QV,(\lambda x.NQ)R]$,
  $E=[\Qr,\QV,N(Q\{R/x\})]$, $\alpha=\lcom$, $\beta=\lbeta$. Clearly, 
  $[\Qr,\QV,(\lambda x.NQ)R]\redto{\lbeta}[\Qr,\QV,N(Q\{R/x\})]$.
\end{varitemize}
$M$ cannot be in the form $\new{c}$, because in that
case $D\equiv E$. This concludes the proof.
\end{proof}

As a simple
corollary of the previous lemma we have the following one--step
confluence property for configurations:
}%
{}%
\begin{proposition}[One-step Confluence]\label{prop:onestepconf}
  Let $C,D,E$ be configurations with  
  $C\predto{\alpha} D$, $C\predto{\beta}E$ and $D\neq E$.
\begin{varenumerate}
\item
  If $\alpha\in\commrul$ and $\beta\in\commrul$, then there is $F$ 
  with $D\predto{\commrul}F$ and $E\predto{\commrul}F$.
\item 
  If $\alpha\in\noncommrul$ and $\beta\in\noncommrul$, then there is $F$ 
  with $D\predto{\noncommrul}F$ and $E\predto{\noncommrul}F$.
\item 
  If $\alpha\in\commrul$ and $\beta\in\noncommrul$, then either 
  $D\predto{\noncommrul}E$ or there 
  is $F$ with $D\predto{\noncommrul} F$ and $E\predto{\commrul}F$.
\end{varenumerate}
\end{proposition}

The fact a strong confluence result like Proposition
\ref{prop:onestepconf} holds here is a consequence of having
adopted the so-called \emph{surface reduction}: it is not 
possible to reduce inside a subterm in the form $!M$ and, as 
a consequence, it is not possible to erase a diverging term.
This has been already pointed out by Simpson~\cite{Simpson05}.

Even in absence of types, we cannot build an infinite sequence of commuting reductions:
\begin{lemma}\label{lemma:noinfcom}
  The relation $\predto{\commrul}$ is strongly
  normalizing. In other words, there cannot be
  any infinite sequence $C_1\predto{\commrul} C_2\predto{\commrul} C_3\predto{\commrul}\ldots$. 
\end{lemma}
\condinc{
 \begin{proof}
Define the size $|M|$ of a term $M$ as the number of symbols in it.
Moreover, define the abstraction size $|M|_\lambda$ of $M$ as the
sum over all subterms of $M$ in the form $\lambda\pi.N$, of
$|N|$. Clearly $|M|_\lambda\leq |M|^2$. Moreover, 
if $[\Qr,\QV,M]\predto{\commrul}[\Qr,\QV,M']$, then $|M'|=|M|$ but $|M'|_\lambda\gt|M|_\lambda$.
This concludes the proof.
\end{proof}
}
{}

The following definition is useful when talking about
reduction lengths, and takes into account both commuting
and non-commuting reductions:

 \begin{definition}
 Let $C_1,\ldots,C_n$ be a sequence of \condinc{(pre)}{}configurations such that
 $C_1\redto\ldots\redto C_n$. The sequence is called an \emph{$m$-sequence
 of length $n$ from $C_1$ to $C_n$} iff $m$ is a natural number and there is $A\subseteq\{1,\ldots,n-1\}$
 with $|A|=m$ and $C_i\ncoredto C_{i+1}$ iff $i\in A$. If there is
 a $m$-sequence of length $n$ from $C$ to $C'$, we will write
 \condinc{$C\mnseq{m}{n}C'$ or simply}{} $C\mseq{m}C'$.
 \end{definition}

\condinc{

\begin{lemma}\label{lemma:commequi}
Let $C,C',D$ be preconfigurations with
$C\equiv C'$ and $C\predto{\alpha}D$. Then there
is $D'\equiv D$ with $C'\predto{\alpha}D'$.
\end{lemma}
\begin{proof}
Let $C=[\Qr,\QV,M]$. We go by induction on $M$.
\end{proof}

\begin{proposition}\label{prop:oscgen}
  Let $C,C',D,D'$ be preconfigurations with 
  $C\predto{\alpha} D$, $C'\predto{\beta}D'$. Then exactly
  one of the following conditions hold:
  \begin{varenumerate}
    \item
      There are $E,E'$ with $E\equiv E'$ such that
      $D\predto{\beta}E$, $D'\predto{\alpha}E'$.
    \item
      $\alpha\in\commrul$, $\beta\in\noncommrul$ and
      there is $E$ with $E\equiv D'$
      such that $D\predto{\beta} E$
    \item
      $\alpha\in\noncommrul$, $\beta\in\commrul$ and
      there is $E$ with $E\equiv D$
      such that $D'\predto{\alpha} E$.
  \end{varenumerate}
\end{proposition}
\begin{proof}
  An easy corollary of Lemma~\ref{lemma:onestepconf}
  and Lemma~\ref{lemma:commequi}.
\end{proof}

}
{}


This way we can generalize \condinc{ Lemma~\ref{lemma:onestepconf}} {
  Proposition~\ref{prop:onestepconf}} to another one talking about
reduction sequences of arbitrary length:

\condinc{
\begin{proposition}\label{OLDprop:manystepconf}
Let $C,D,D'$ be preconfigurations with $C\mseq{m}D$ and
$C\mseq{m'}D'$. Then, there are preconfigurations
$E,E'$ with $E\equiv E'$, $D\mseq{n}E$ and $D'\mseq{n'}E'$ 
with $n\leq m'$, $n'\leq m$ and $n+m=n'+m'$.
\end{proposition}
\begin{proof}
  We prove the following, stronger statement: suppose there are
  $C,C',D,D'$ with $C\equiv C'$, a $m$-sequence of length $l$ from $C$
  to $D$ and an $m'$-sequence of length $l'$ from $C'$ to $D'$. Then,
  there are a preconfiguration $E,E'$ with $E\equiv E'$, a
  $n$-sequence of length $k$ from $D$ to $E$ and $n'$-sequence of
  length $k'$ from $D'$ to $E'$ with $n\leq m'$, $n'\leq m$, $k\leq
  l'$, $k'\leq l$ and $n+m=n'+m'$.  We go by induction on $l+l'$. If
  $l+l'=0$, then $C=D$, $C'=D'$, $E=D$, $E'=D'$ and all the involved
  natural numbers are $0$.  If $l=0$, then $D=C$, $E'=D'$ and $E$ can
  be obtained applying $l'$ times
  Lemma~\ref{lemma:commequi}. Similarly when $l'=0$. So, we can assume
  $l,l'\gt 0$.  There are $G,G'$, two integers $h,h'\leq 1$ with
  $C\predto{\alpha} G'$ and $C'\predto{\beta} G'$, an $(m-h)$-sequence
  of length $l-1$ from $G$ to $D$ and an $(m'-h')$-sequence of length
  $l'-1$ from $G'$ to $D'$ We can distinguish three cases, depending
  on the outcome of Proposition~\ref{prop:oscgen}:
\begin{varitemize}
\item
  There are $H,H'$ with $G\predto{\beta} H$ and $G'\predto{\alpha} H'$.
  By applying several times the induction hypothesis, we end up with
  the following diagram

  \centerline{
  \xymatrix@C=10pt
  {
    &  &  & C\ar@{--}[r]\ar[ld]_{h,1} & C'\ar[rd]^{h',1} &  &    & \\
    & G\ar@{--}[r]\ar[ld]_{m-h,l-1} & G\ar[rd]_{h',1} &  &  & G'\ar@{--}[r]\ar[ld]^{h,1} & G'\ar[rd]^{m'-h',l'-1} & \\
    D\ar[rd]_{q,t} & & & H\ar@{--}[r]\ar[ld]_{u,v} & H'\ar[rd]^{u',v'} & & & D'\ar[ld]^{q',t'} \\
    & J\ar@{--}[r]\ar[rd]_{a,b} & K\ar[rd]^{w,z} & & & K'\ar@{--}[r]\ar[ld]_{w',z'} & J'\ar[ld]^{a',b'} & \\
    & & E\ar@{--}[r] & L\ar@{--}[r] & L'\ar@{--}[r] & E & &
  }}
  together with the equations:
  $$
  \begin{array}{llll}
    q\leq h' & q'\leq h & w\leq u' & a=w \\
    t\leq 1 & t'\leq 1 & z\leq  v' & b\leq z \\
    u\leq m-h & u'\leq m' & w'\leq u & a'=w' \\
    v\leq l-1 & v'\leq l'-1 & z'\leq v & b'\leq z'
  \end{array}
  $$
  and
  $$
  \begin{array}{ccc}
    m-h+q=u+h' & h+u'=m'-h'+q' & w+u = w'+u'
  \end{array} 
  $$
  from which
  \begin{eqnarray*}
    q+a&\leq& h'+w\leq h'+u'\leq h'+m'-h'=m'\\
    b+t&\leq& z+1\leq v'+1\leq l'-1+1=l'\\
    q'+a'&\leq&h+w'\leq h+u\leq h+m-h=m\\
    b'+t'&\leq&z'+1\leq v+1\leq l-1+1=l\\
    q+a+m&=&h+h'+u+w=h+h'+u'+w'=m'+a'+q'
  \end{eqnarray*}
  So we can just put $n=a$, $n'=q'+a'$, $k=t+b$, $k'=t'+b'$.
\item 
  $\alpha\in\commrul$, $\beta\in\noncommrul$ and there is 
  $H$ with $G\equiv H$ and $G'\predto{\beta} H$.
  By applying several times the induction hypothesis, we end up with
  the following diagram:

  \centerline{
  \xymatrix@C=10pt
  {
      &  & C\ar@{--}[r]\ar[ld]_{h,1} & C'\ar[rd]^{0,1} &  &    & \\
     G\ar@{--}[r]\ar[dd]_{m-h,l-1} & G\ar@{--}[rd] &  &  & G'\ar@{--}[r]\ar[ld]^{h,1} & G'\ar[rd]^{m',l'-1} & \\
     & & H\ar@{--}[r]\ar[ld]_{u,v} & H\ar[rd]^{u',v'} & & & D'\ar[ld]^{q',t'} \\
     D\ar@{--}[r]\ar[rd]_{a,b} & K\ar[rd]^{w,z} & & & J'\ar@{--}[r]\ar[ld]_{w',z'} & K'\ar[ld]^{a',b'} & \\
     & E\ar@{--}[r] & L\ar@{--}[r] & L'\ar@{--}[r] & E & &
  }}
  together with the equations:
  $$
  \begin{array}{llll}
            & q'\leq h & w\leq u' & a=w \\
            & t'\leq 1 & z\leq  v' & b\leq z \\
    u\leq m-h & u'\leq m'-h' & w'\leq u & a'=w' \\
    v\leq l-1 & v'\leq l'-1 & z'\leq v & b'\leq z'
  \end{array}
  $$
  and
  $$
  \begin{array}{ccc}
    u=m-h & h+u'=m'+q' & w+u = w'+u'
  \end{array} 
  $$
  from which
  \begin{eqnarray*}
    a&\leq& w\leq u'\leq m'\\
    b&\leq& z\leq v'\leq l'-1\leq l'\\
    q'+a'&\leq&h+w'\leq h+u\leq h+m-h=m\\
    b'+t'&\leq&z'+1\leq v+1\leq l-1+1=l\\
    a+m&=&w+u+h=w'+u'+h=w'm'+q'=a'+m'+q'
  \end{eqnarray*}
  So, we can just put $n=a$, $n'=a'+q'$, $k=b$, $k'=t'+b'$.
\item
  The last case is similar to the previous one.
\end{varitemize}
This concludes the proof.
\end{proof}
}
{}

As a direct consequence of the previous proposition we have:

\begin{proposition}\label{prop:manystepconf}
Let $C,D,E$ be configurations with $C\mseq{m}D$ and
$C\mseq{n}E$. Then, there is a configuration
$F$ with $D\mseq{p}F$ and $E\mseq{q}F$,
where $p\leq n$, $q\leq m$ and $p+m=q+n$.
\end{proposition}

Finally, we can prove the main result of this section:
\begin{theorem}\label{teo:weak-strong}
 \condinc{A configuration}{} $C$ is strongly normalizing iff $C$ is weakly normalizing.
\end{theorem}
\begin{proof}
Strong normalization implies weak normalization. Suppose, by way
of contradiction, that $C$ is weakly normalizing but not strongly
normalizing. This implies there is a configuration $D$ in normal
form and an $m$-sequence from $C$ to $D$. Since $C$ is not
strongly normalizing, there is an infinite sequence
$C=C_1,C_2,C_3,\ldots$ with $C_1\redto C_2\redto C_3\redto\ldots$
From this infinite sequence, we can extract an $m+1$-sequence,
due to Lemma~\ref{lemma:noinfcom}. Applying Proposition~\ref{prop:manystepconf},
we get a configuration $F$ and a $1$-sequence from $D$ to $F$. 
However, such a $1$-sequence cannot exist, because $D$ is normal.
\end{proof}
\section{Standardizing Computations}
One of the main interesting properties of the \qcalc\ is the
capability of performing computational steps in the following order:
\begin{varitemize}
\item
First perform classical reductions.
\item
Secondly, perform reductions that build the underlying quantum register.
\item
Finally, perform quantum reductions.
\end{varitemize}
We distinguish three particular subsets of $\redrul$, namely
$\quantrul=\{\Uq,\qbeta\}$, $\noncrul=\quantrul\cup\{\nw\}$,
and $\classrul=\redrul-\noncrul$.
Let $C\predto{\alpha} C'$ and let $M$ be the relevant redex in $C$;
if $\alpha\in\quantrul$ the redex $M$ is called \textit{quantum}, if
$\alpha\in \classrul$ the redex $M$ is called \textit{classical}.
\begin{definition} 
  A configuration $C$ is called \textit{non classical} if
  $\alpha\in\noncrul$ whenever $C\predto{\alpha} C'$. Let $\Ncl$ be
  the set of non classical configurations.
  A configuration $C$ is called \textit{essentially quantum} if
  $\alpha\in\quantrul$ whenever $C\predto{\alpha} C'$. Let $\Eqt$ be
  the set of essentially quantum configurations.
\end{definition}
Before claiming the standardization theorem, we need the following definition: 
\begin{definition} A $\CQ$ computation starting with a configuration
  $C$ is a computation $\{C_i\}_{i\lt\varphi}$ such that $C_0=C$ and:
  \begin{varenumerate}
  \item for
  every $0\lt i\lt\varphi-1$, if $C_{i-1}\predto{\noncrul} C_i$ then
  $C_i\predto{\noncrul} C_{i+1}$;
\item for
  every $0\lt i\lt\varphi-1$, if $C_{i-1}\predto{\quantrul} C_i$ then
  $C_i\predto{\quantrul} C_{i+1}$.
  \end{varenumerate}
 \end{definition}

More informally, a $\CQ$ computation is a computation when $\nw$ reductions are
always performed after classical reductions and before quantum reductions.

$\Ncl$ is closed under $\nw$ reduction, while $\Eqt$ is closed under quantum reduction:
\begin{lemma} \label{lemma:NCLclosure}
If $[\Qr,\QV, M] \in \Ncl$ and $[\Qr,\QV, M]
  \predto{\nw}[\Qr',\QV', M']$ then $[\Qr',\QV', M']\in \Ncl$.
\end{lemma}
\condinc{
\begin{proof}
  \newcommand{\Cx}[1]{\mathbf{C}[#1]} %
  Let us denote with $\Cx{\ }$ a generic context that does not
  contain classical redexes, and let $\new{c}$ be the reduced redex in
  $M$.

  The proof proceeds by cases on the structure of $M$.

there are several  case:
    \begin{enumerate}
    \item $M \equiv \new{c}$ and $M'\equiv q$.\\
      Observe that $M'$ is in normal form and conclude
    \item $M\equiv \Cx{L(\new{c})}$ and $M'\equiv
      \Cx{Lq}$
      Observe that $L$ cannot be $\lambda x.R$ because $M\in \Ncl$ and
      therefore no classical redexes
      can be generated.
    \item in the following cases it is immediate to observe that the
      reduction does not generate classical redexes:
      \begin{enumerate}
      \item $M\equiv \Cx{(\new{c})L}$ and $M'\equiv
        \Cx{qL}$;
      \item $M\equiv \Cx{\lambda! x .(\new{c})}$ and
        $M' \equiv \Cx{\lambda! x.q}$.
      \item $M\equiv \Cx{<N_1,\ldots,N_{k-1},(\new{c}),N_{k+1}\ldots,N_w>}$
        \\ and \\
        $M'\equiv \Cx{<N_1,\ldots,N_{k-1}, q ,N_{k+1}\ldots,N_w>}$
      \item $M\equiv\Cx{\pnew{(\new{c})}}$ and $M'\equiv \Cx{\pnew{q}}$
      \item $M\equiv\Cx{\qnew{(\new{c})}}$ and $M'\equiv \Cx{\qnew{q}}$
      \end{enumerate}
    \end{enumerate}
\end{proof}
}
{}

\begin{lemma}\label{lemma:EQTclosure}
If $[\Qr,\QV, M] \in \Eqt$ and $[\Qr,\QV, M]
  \predto{\quantrul}[\Qr',\QV', M']$ then $[\Qr',\QV', M']\in \Eqt$.
\end{lemma}
\condinc{
\begin{proof} 
  \newcommand{\Cx}[1]{\mathbf{C}[#1]} 
  Let us denote with $\Cx{\ }$ a generic context that does not
  contain classical redexes. Let $P \equiv
  \lambda<x_1,\ldots,x_n>.N <r_1,\ldots,r_n>$ and $R\equiv
  U<r_1,\ldots,r_n>$ be two quantum redexes and let $N'\equiv
  N\{r_{1}/x_{1},\ldots,r_{n}/x_{n}\}$

  The proof proceeds by case on the shape of the reduced redex in $M$
  (and by subcases on the structure of $M$).
  \begin{description}
  \item[case 1: the reduced redex is $P$]\mbox{}\\
    Cause commutative reductions, it is impossible that $M$ is
    $\Cx{P L}$. \\
     Let us examine all the possible cases:
    \begin{enumerate}
    \item $M\equiv P$ and
      $M'\equiv N'$.\\
      It is trivial to observe that $M'$ cannot contain classical
      redexes.
    \item $M\equiv \Cx{LP}$ and
      $M'\equiv \Cx{LN'}\equiv\Cx{(L)N\{r_{1}/x_{1},\ldots,r_{n}/x_{n}\}}$:\\
      The reduction could (hypothetically) create a (new) classical
      redex in $M'\equiv \Cx{(L)N'}$ iff :
      \begin{enumerate}
      \item $L\equiv \lambda !z. L'$ and $N' \equiv !N''$: impossible
        because in this case $P$ should have the shape\\
        $\lambda<x_1,\ldots,x_n>.!N''' <r_1,\ldots,r_n>$, but this
        term is not well formed;
      \item $L\equiv \lambda <z_1,\ldots,z_r>. L'$ and $N'\equiv
        <N'_1,\ldots,N'_r>$ : in this case $M$ should be:\\ $\Cx{(\lambda
          <z_1,\ldots,z_r>. L')(\lambda<x_1,\ldots,x_n>.N
          <r_1,\ldots,r_n>)}$, but this is impossible because in this
        case $M$ has a commutative redex;
      \item $L\equiv \lambda <z_1,\ldots,z_r>. L'$ and $N'\equiv
        (\lambda\pi'.N')N''$: impossible because $M$ should be\\
        $\Cx{(\lambda <z_1,\ldots,z_r>. L')(\lambda<x_1,\ldots,x_n>.N
          <r_1,\ldots,r_n>)}$ and $M$ should have a commutative redex.
      \item $N'\equiv (\lambda \pi' .N')N''$: 
      \end{enumerate}
    \item in the following cases:
      \begin{enumerate}
      \item $M\equiv \Cx{\lambda\pi.P}$ and
        $M' \equiv \Cx{\lambda \pi.N'}$.
      \item $M\equiv \Cx{\lambda! x .P}$ and
        $M' \equiv \Cx{\lambda! x.N'}$.
      \item $M\equiv \Cx{<N_1,\ldots,N_{k-1},P,N_{k+1}\ldots,N_w>}$
        \\ and \\
        $M'\equiv \Cx{<N_1,\ldots,N_{k-1}, N' ,N_{k+1}\ldots,N_w>}$.
      \end{enumerate}
      by means of (1) it is immediate to observe that  the reduction does not
      generate classical redex in $M'$.
    \item $M\equiv\Cx{\pnew{P}}$ and $M'\equiv \Cx{\pnew{N'}}$\\
      It is immediate to observe that the reduction does not generate
      new classical redexes, in fact by linearity, $N'$ cannot be a
      neither $0$ nor $1$.
    \item  $M\equiv\Cx{\qnew{P}}$ and $M'\equiv \Cx{\qnew{N'}}$\\
      as for the previous case.
    \end{enumerate}
  \item[case 2: the reduced redex is $R$] \mbox{}\\
    there are several case:
    \begin{enumerate}
    \item $M \equiv R$ and $M'\equiv <r_1,\ldots,r_n>$.\\
      $M'$ is in normal form.
    \item $M\equiv \Cx{L(U<r_1,\ldots,r_n>)}$ and $M'\equiv
      \Cx{L<r_1,\ldots,r_n>}$\\
      Observe that $L$ cannot be $\lambda x.R$ because $M\in \Eqt$ and
      therefore no classical redexes
      can be generated.\\
    \item in the following cases it is immediate to observe that the
      reduction does not generate classical redexes:
      \begin{enumerate}
      \item $M\equiv \Cx{(U<r_1,\ldots,r_n>)L}$ and $M'\equiv
        \Cx{<r_1,\ldots,r_n>L}$;
      \item $M\equiv \Cx{\lambda! x .(U<r_1,\ldots,r_n>)}$ and
        $M' \equiv \Cx{\lambda! x.<r_1,\ldots,r_n>}$.
      \item $M\equiv \Cx{<N_1,\ldots,N_{k-1},(U<r_1,\ldots,r_n>),N_{k+1}\ldots,N_w>}$
        \\ and \\
        $M'\equiv \Cx{<N_1,\ldots,N_{k-1}, <r_1,\ldots,r_n> ,N_{k+1}\ldots,N_w>}$
      \item $M\equiv\Cx{\pnew{(U<r_1,\ldots,r_n>)}}$ and $M'\equiv \Cx{\pnew{<r_1,\ldots,r_n>}}$
      \item $M\equiv\Cx{\qnew{(U<r_1,\ldots,r_n>)}}$ and $M'\equiv \Cx{\qnew{<r_1,\ldots,r_n>}}$
      \end{enumerate}
    \end{enumerate}
\end{description}
\end{proof}
}
{}
This way we are able to state the  Standardization Theorem.
\begin{theorem}[Standardization]
  For every computation $\{C_i\}_{i\lt\varphi}$ such that $\varphi\in\NN$
  there is a $\CQ$ computation $\{C'_i\}_{i\lt\xi}$ such that $C_0=C'_0$ and 
   $C_{\varphi-1}=C'_{\xi-1}$.
\end{theorem}
\begin{proof}
We will build a $\CQ$ computation in three steps:
\begin{varenumerate}
\item 
  Let us start to reduce $C'_0=C_0$ \textit{by using}
  $\classrul$ reductions as much as possible. By 
  Theorem~\ref{teo:weak-strong} we must obtain a finite reduction sequence
  $C'_0\predto{\classrul}\ldots\predto{\classrul}C'_k$
  s.t. $0\leq k<\varphi$ and no 
  $\classrul$ reductions are applicable to $C'_k$
\item
  Reduce $C'_k$ \textit{by using} $\nw$ reductions  
  as much as possible. By 
  Theorem~\ref{teo:weak-strong} we must obtain a finite reduction sequence
  $C'_k\predto{\nw}\ldots\predto{\nw}C'_j$ s.t. $k\leq j\lt\varphi$ and no
  $\nw$ reductions are applicable to $C'_j$.  Note that by
  Lemma~\ref{lemma:NCLclosure} such reduction steps cannot generate
  classical redexes and in particular no classical redex can
  appear in $C'_j$.
\item 
  Reduce $C'_j$ \textit{by using} $\quantrul$ reductions 
  as much as possible. By
  Theorem~\ref{teo:weak-strong} we must obtain a finite reduction sequence
  $C'_j\predto{\quantrul}\ldots\predto{\quantrul}\cdots C'_{m}$
  such that $j\leq m<\varphi$ and no $\quantrul$ reductions are applicable to
  $C'_{m}$.  Note that by Lemma~\ref{lemma:EQTclosure} such 
  reduction steps cannot generate neither $\classrul$ redexes nor $\nw$
  redexes and in particular neither $\classrul$ nor $\nw$
  reductions are applicable to $C'_{m}$. Therefore $C'_{m}$\textit{ is
  in normal form}.
\end{varenumerate}
The reduction sequence $\{C'_i\}_{i\lt m+1}$ is such that
$C'_0\predto{\classrul}\ldots\predto{\classrul}
C'_k\predto{\nw}\ldots\predto{\nw}
C'_j\predto{\quantrul}\ldots\predto{\quantrul} C'_{m}$ is a
$\CQ$ computation.
By Proposition~\ref{prop:manystepconf} we observe that
$C_{\varphi-1}= C'_{m}$.
\end{proof}

The \emph{intuition} behind a $\CQ$ computation is the following:
the first phase of the computation is responsible for the construction
of a $\lambda$--term (abstractly) representing a quantum
circuit and does not touch the underlying quantum register. 
The second phase builds the quantum register without introducing
any superposition. The third phase corresponds to proper quantum 
computation (unitary operators are applied to the quantum register,
possibly introducing superposition).
This intuition will become a technical recipe in order to prove a side
of the equivalence between \qcalc\ and quantum circuit families
formalism (see Section~\ref{subsect:qcal-circ}).

\section{Expressive Power}
In this section we study the expressive power of the \qcalc,
showing that it is equivalent to finitely generated
quantum circuit families, and consequently (via the result of 
Ozawa and Nishimura~\cite{NiOz02}) we have the equivalence with 
quantum Turing machines as defined by
Bernstain and Vazirani\cite{BerVa97}. The fact the considered class
of circuit families only contains finitely generated ones is not
an accident: if we want to represent an entire family by one
single lambda term (which is, by definition, a finite object) we
must restrict to families which are generated by a discrete set
of gates.

\subsection{Encoding Quantum Circuits Families}
In this Section we  will show that each (finitely generated) quantum
circuit family can be captured by a \textit{quantum relevant} term. 
\condinc{\subsubsection{Classical Strength of the \qcalc.}}
        {\paragraph{Classical Strength of the \qcalc.}}
The classical fragment of the \qcalc\ has the expressive power
of pure, untyped lambda calculus.
\condinc{
\begin{lemma}
If $[\Qr,\QV,M]\predto{\scrul}[\Qr',\QV',M']$,
then $\Qr=\Qr'$ and $\QV=\QV'$.
\end{lemma}
\paragraph{Natural Numbers}
Natural numbers are encoded as follows:
\begin{eqnarray*}
  \nat{0}&=&\lambda !x.\lambda !y.y\\
  \forall n.\nat{n+1}&=&\lambda !x.\lambda !y.x!\nat{n}
\end{eqnarray*}
This way, we can compute the successor and the predecessor
of a natural number as follows:
\begin{eqnarray*}
  \tsuc&=&\lambda !z.\lambda !x.\lambda !y.x!z\\
  \pred&=&\lambda !z.z!(\lambda !x.x)!\nat{0}
\end{eqnarray*}
Indeed:
\begin{eqnarray*}
  \tsuc\ssp!\nat{n}&\predto{\classrul}&\lambda !x.\lambda !y.x!\nat{n}=\nat{n+1};\\
  \pred\ssp!\nat{0}&\predto{\classrul}&\nat{0}!(\lambda !x.x)!\nat{0}\predto{\classrul}\nat{0};\\
  \pred\ssp!\nat{n+1}&\predto{\classrul}&\nat{n+1}!(\lambda !x.x)!\nat{0}\predto{\classrul}(\lambda !x.x)!\nat{n}\\
     &\predto{\classrul}&\nat{n}
\end{eqnarray*}
\paragraph{Lists.}
Given a sequence $M_1,\ldots,M_n$ of terms, 
we can build a term $\blist{M_1,\ldots,M_n}$ encoding
the sequence as follows, by induction on $n$:
\begin{eqnarray*}
  \blist{}&=&\lambda !x.\lambda !y.y;\\
  \blist{M,M_1\ldots,M_n}&=&\lambda !x.\lambda !y.xM\blist{M_1,\ldots,M_n}.
\end{eqnarray*}
This way we can construct and destruct lists in a principled way:
terms $\cons$ and $\sel$ can be built as follows:
\begin{eqnarray*}
  \cons&=&\lambda z.\lambda w.\lambda !x.\lambda !y.xzw;\\
  \sel&=&\lambda x.\lambda y.\lambda z.xyz.
\end{eqnarray*}
They behave as follows on lists:
\begin{eqnarray*}
  \cons\ssp M \blist{M_1,\ldots,M_n}&\rpredto{\classrul}&\blist{M,M_1,\ldots,M_n}\\
  \sel\ssp\blist{}!N!L&\rpredto{\classrul}&L\\
  \sel\ssp\blist{M,M_1,\ldots,M_n}!N!L&\rpredto{\classrul}&NM\blist{M_1,\ldots,M_n}
\end{eqnarray*}
By exploiting $\cons$ and $\sel$, we can build more advanced
constructors and destructors: for every natural number $n$ there
are terms $\append{n}$ and $\extract{n}$ behaving as follows:
\begin{eqnarray*}
  \append{n}\blist{N_1,\ldots,N_m}M_1,\ldots,M_n&\rpredto{\scrul}&\blist{M_1,\ldots,M_n,N_1,\ldots,N_m}\\
  \forall m\leq n.\extract{n} M\blist{N_1,\ldots,N_m}&\rpredto{\scrul}&M\blist{}N_m N_{m-1}\ldots N_1\\
  \forall m\gt n.\extract{n} M\blist{N_1,\ldots N_m}&\rpredto{\scrul}&M\blist{N_{n+1}\ldots N_m}N_n N_{n-1}\ldots N_1
\end{eqnarray*}
Terms $\append{n}$ can be built by induction on $n$:
 \begin{eqnarray*}
   \append{0}&=&\lambda x.x\\
   \append{n+1}&=&\lambda x.\lambda y_1.\ldots.\lambda y_{n+1}.\cons\ssp y_{n+1} (\append{n} x y_{1}\ldots y_{n})
 \end{eqnarray*}
Similarly, terms $\extract{n}$ can be built inductively:
\begin{eqnarray*}
  \extract{0}&=&\lambda x.\lambda y.xy\\
  \extract{n+1}&=&\lambda x.\lambda y.(\sel y !(\lambda z.\lambda w.\lambda v.\extract{n} v w z)!(\lambda z.z\blist{}))x
\end{eqnarray*}
Indeed:
\begin{eqnarray*}
  \extract{0}M\blist{N_1,\ldots N_m}&\rpredto{\scrul}&M\blist{N_1,\ldots,N_m}\\
  \extract{n+1} M\blist{}&\rpredto{\scrul}& M\blist{}\\
  \forall m\leq n.\extract{n+1}M\blist{N,N_1\ldots N_m}&\rpredto{\scrul}&\extract{n}M\blist{N_1,\ldots,N_m}N\\
  &\rpredto{\scrul}&M\blist{}N_m\ldots N_1 N\\
  \forall m\gt n.\extract{n+1}M\blist{N,N_1\ldots N_m}&\rpredto{\scrul}&\extract{n}M\blist{N_1,\ldots,N_m}N\\
  &\rpredto{\scrul}&M\blist{N_{n+1}\ldots N_m}N_n\ldots N_1 N\\
\end{eqnarray*}
\paragraph{Recursion and Iteration.}
We now need a term for iteration: $\rec$ is defined as $\recaux!\recaux$, where
$$
\recaux=\lambda !x.\lambda !y.y!((x!x)!y).
$$
For each term $M$,
\begin{eqnarray*}
 \rec!M&=&(\recaux!\recaux)!M\predto{\scrul}(\lambda !y.y!((\recaux !\recaux)!y))!M\\
   &\predto{\scrul}&M!((\recaux !\recaux)!M))=M!(\rec!M)
 \end{eqnarray*}
This will help us in encodings algorithms via recursion. If one wants
to iterate a given function over natural numbers, there is $\iternat=\rec!\iternataux$, where
$$
\iternataux=\lambda!x.\lambda !y.\lambda !w.\lambda !z.y!(\lambda !v.w!(x!v!w !z)!v)!z
$$
Indeed:
\begin{eqnarray*}
  \iternat\ssp\nat{0}!M!N&\rpredto{\scrul}&\iternataux!(\iternat)\nat{0}!M!N\\
  &\rpredto{\scrul}&\nat{0}!(\lambda !v.M (\iternat v !M !N)!v)!N\\
  &\rpredto{\scrul}& N\\
  \iternat\ssp\nat{n+1}!M!N&\rpredto{\scrul}&\nat{n+1}!(\lambda !v.M !(\iternat !v !M !N)!v)!N\\
  &\rpredto{\scrul}&(\lambda !v.M (\iternat !v !M !N)!v)!\nat{n}\\
  &\rpredto{\scrul}&M (\iternat\ssp!\nat{n} !M !N)!\nat{n}
\end{eqnarray*}
\begin{definition}
A function $f:\NN^n\rightarrow\NN$ is representable iff there is a term $M_f$
such that:
\begin{varitemize} 
  \item
    Whenever $M_f\nat{m_1}\ldots\nat{m_n}$ has a normal
    form $N$ (with respect to $\rpredto{\classrul}$), then $N=\nat{m}$
    for some natural number $m$.
  \item
    $M_f\nat{m_1}\ldots\nat{m_n}\rpredto{\classrul}\nat{m}$ iff
    $f(m_1,\ldots,m_n)$ is defined and equal to $m$.
\end{varitemize}
\end{definition}
\begin{proposition}
The class of representable functions coincides with the class of
partial recursive functions (on natural numbers).
\end{proposition}
Iteration is available on lists, too. Let $\iterlist=\rec!\iterlistaux$, where
$$
\iterlistaux=\lambda!x.\lambda y.\lambda !w.\lambda !z.y!(\lambda v.\lambda u.w(xu!w!z)v)!z
$$
Indeed:
\begin{eqnarray*}
  \iterlist\ssp\blist{}!M!N&\rpredto{\classrul}&\iterlistaux!(\iterlist)\blist{}!M!N\\
  &\rpredto{\classrul}&\blist{}!(\lambda v.\lambda u.M(\iterlist\ssp u!M!N)v)!N\\
  &\rpredto{\classrul}& N\\
  \iterlist\ssp\blist{L,L_1,\ldots,L_n}!M!N&\rpredto{\classrul}&
     \blist{L,L_1,\ldots,L_n}!(\lambda v.\lambda u.M(\iterlist\ssp u!M!N)v)!N\\
  &\rpredto{\classrul}&(\lambda v.\lambda u.M(\iterlist\ssp u!M!N)v)L\blist{L_1,\ldots,L_n}\\
  &\rpredto{\classrul}& M (\iterlist\blist{L_1,\ldots,L_n}!M!N)L
\end{eqnarray*}
}
{For example, we can easily
encode full, higher-order recursion as the term $\recaux!\recaux$, where
$$
\recaux=\lambda !x.\lambda !y.y!((x!x)!y).
$$
Indeed, for each term $M$ (with a slight abuse of notation\footnote{the rewriting
relation $\redto$ is only defined on configurations.}),
\begin{eqnarray*}
\rec!M&=&(\recaux!\recaux)!M\\
      &\predto{\classrul}&(\lambda !y.y!((\recaux !\recaux)!y))!M\\
      &\predto{\classrul}&M!((\recaux !\recaux)!M))\\
      &=&M!(\rec!M)
\end{eqnarray*}
Natural numbers can be encoded as follows:
 Natural numbers are encoded as follows:
 \begin{eqnarray*}
   \nat{0}&=&\lambda !x.\lambda !y.y\\
   \forall n\in\NN.\nat{n+1}&=&\lambda !x.\lambda !y.x\nat{n}
 \end{eqnarray*}
Given a sequence $M_1,\ldots,M_n$ of terms, 
we can build a term $\blist{M_1,\ldots,M_n}$ encoding
the sequence as follows, by induction on $n$:
\begin{eqnarray*}
 \blist{}&=&\lambda !x.\lambda !y.y;\\
 \blist{M,M_1\ldots,M_n}&=&\lambda !x.\lambda !y.xM\blist{M_1,\ldots,M_n}.
\end{eqnarray*}
This way we can construct and destruct lists in a principled way:
terms $\cons$ and $\sel$ can be built as follows:
\begin{eqnarray*}
   \cons&=&\lambda z.\lambda w.\lambda !x.\lambda !y.xzw;\\
   \sel&=&\lambda x.\lambda y.\lambda z.xyz.
\end{eqnarray*}
They behave as follows on lists:
\begin{eqnarray*}
 \cons\ssp M \blist{M_1,\ldots,M_n}&\rpredto{\classrul}&\blist{M,M_1,\ldots,M_n}\\
 \sel\ssp\blist{}!N!L&\rpredto{\classrul}&L\\
 \sel\ssp\blist{M,M_1,\ldots,M_n}!N!L&\rpredto{\classrul}&NM\blist{M_1,\ldots,M_n}
\end{eqnarray*}
}
\condinc{\subsubsection{Quantum Relevant Terms.}}
        {\paragraph{Quantum Relevant Terms.}}
\begin{definition}
  Let $\mathscr{S}$ be any subset of $\redrul$.
  The expression $[\Qr, M]\Downarrow_{\mathscr{S}} [\Qr', M']$ 
  means that $[\Qr,M]\predto{\mathscr{S}}[\Qr', M']$ and
  $[\Qr', M']$ is in normal form with respect to the relation
  $\predto{\mathscr{S}}$. $[\Qr, M]\Downarrow [\Qr', M']$ stands for 
  $[\Qr, M]\Downarrow_\redrul [\Qr', M']$
\end{definition}

Confluence and the equivalence between weakly normalizing and strongly normalizing 
configurations authorize the following definition:
\begin{definition}
  A term $M$ is called \textit{quantum relevant} (shortly
  \textit{q--rel}) if it is well formed and for each list 
  $![!c_1,...,!c_n]$ there are a quantum register $\Qr$ and 
  a natural number $m$ such that
  $[1,M![!c_1,...,!c_n]]\Downarrow[\Qr,[r_1,\ldots,r_m]]$.
\end{definition}
In other words, a quantum relevant term is the analogue 
of a pure $\lambda$-term representing a function on natural numbers.
\begin{remark}
  It is immediate to observe that the class of \textit{q--rel} terms
  in not recursively enumerable.
\end{remark}
\condinc{\subsubsection{Circuits.}}
        {\paragraph{Circuits.}}
An \emph{$n$-qubit gate} (or, simply, a \emph{qubit gate}) is a unitary operator
$U:\CC^{2^{n}}\rightarrow\CC^{2^{n}}$, while
a $\Vr$-qubit gate (where $\Vr$ is a qvs) is a unitary
operator $G:\Hs{\Vr}\rightarrow\Hs{\Vr}$. We here work 
with computable operators only. If $\mathcal{G}$ is a set of
qubit gates, a \emph{$\Lambda$-circuit} $K$ based on $\mathcal{G}$
is a sequence
$$
U_1,r_1^1,\ldots,r_{n_1}^1,\ldots,U_m,r_1^m,\ldots,r_{n_m}^m
$$
where, for every $1\leq i\leq m$:
\begin{varitemize}
\item
  $U_i$ is an $n_i$-qubit gate in $\mathcal{G}$;
\item
  $r_1^i,\ldots,r_{n_i}^i$ are distinct quantum variables 
  in $\Lambda$.
\end{varitemize}
The $\Lambda$-gate $U_K$ \emph{determined by} a $\Lambda$-circuit
$$
K=U_1,r_1^1,\ldots,r_{n_1}^1,\ldots,U_m,r_1^m,\ldots,r_{n_m}^m
$$
is the unitary operator
$$
(U_m)_{\langle\langle r_1^m,\ldots,r_{n_m}^m\rangle\rangle}\circ\ldots\circ
(U_1)_{\langle\langle r_1^1,\ldots,r_{n_1}^1\rangle\rangle}.
$$
Let $(\mathbf{K}_i)_{i\lt \omega}$ be an effective enumeration of
quantum circuits.

A \emph{family of circuits} generated by $\mathcal{G}$ is a
triple $(f,g,h)$ where:
\begin{varitemize}
\item
  $f:\NN\rightarrow\NN$ is a computable function;
\item
  $g:\NN\times\NN\rightarrow\NN$ is a computable
  function such that $0\leq g(n,m)\leq n+1$ whenever
  $1\leq m\leq f(n)$;
\item
  $h:\NN\rightarrow\NN$ is a computable function
  such that for every $n\in\NN$, $\mathbf{K}_{h(n)}$ is a $\{r_1,\ldots,r_{f(n)}\}$-circuit
  based on $\mathcal{G}$.
\end{varitemize}
A family of circuits $(f,g,h)$ generated by a finite set
$\mathcal{G}$ is said to be \emph{finitely generated}. 
\condinc{
\subsubsection{The Result.}
The $n$-th elementar permutation of $m$ elements
(where $1\leq n\lt m$) is the function which maps 
$n$ to $n+1$, $n+1$ to $n$ and any other elements 
in the interval $1,\ldots, m$ to itself.
A term $M$ \emph{computes the $n$-th elementary permutation
on lists} iff for every list $[N_1,\ldots,N_m]$ with $m\gt n$,
$M[N_1,\ldots,N_m]\rpredto{\classrul}[N_1,\ldots,N_{n-1},N_{n+1},N_{n},N_{n+2},\ldots,N_m]$.
\begin{lemma}\label{lemma:existperm}
There is a term $M$ such that, for every natural number $n$,
$M\nat{n}$ computes the $n+1$-th elementary permutation on lists.
\end{lemma}
\begin{proof}
$M$ is the term
$$
\lambda !x.\iternat !x
  !(\lambda !y.\lambda !z.\lambda w.\extract{1}(\lambda q.\lambda s.\append{1}(yq)s)w)
  !(\lambda y.\extract{2}(\lambda z.\lambda w.\lambda q.\append{2} zqw)y)
$$
This completes the proof.
\end{proof} 
\begin{lemma}\label{lemma:listonat}
There is a term $M$ such that, for every list
$[!N_1,\ldots,!N_n]$, $M[!N_1,\ldots,!N_n]\rpredto{\classrul}\nat{n}$.
\end{lemma}
\begin{proof}
$M$ is the term
$$
\lambda x.\iterlist x!(\lambda y.\lambda !z.\tsuc y)!(\nat{0})
$$
This completes the proof.
\end{proof} 
\begin{lemma}\label{lemma:listnattoele}
There is a term $M$ such that for every list $[!N_1,\ldots,!N_m]$:
\begin{eqnarray*}
  M!\nat{0}![!N_1,\ldots,!N_m]&\rpredto{\classrul}&!0\\
  \forall 1\leq n\leq m.M!\nat{n}![!N_1,\ldots,!N_m]&\rpredto{\classrul}&!N_n\\
  M!\nat{m+1}![!N_1,\ldots,!N_m]&\rpredto{\classrul}&!1
\end{eqnarray*}  
\end{lemma}
\begin{proof}
$M$ is the term
$$
\lambda !x.\lambda !y.(\iterlist y
                      !(\lambda z.\lambda !w.\lambda !q.(q!(\lambda !s.\lambda r.(s!L_{\geq 2}!L_{=1})r)!L_{=0})z)
                       !(\lambda !z.z!(\lambda !w.!1)!0)
                      )!x
$$
where
\begin{eqnarray*}
L_{=0}&=&\lambda t.t!\nat{0}\\
L_{=1}&=&\lambda t.(\lambda !u.!w)(t!\nat{0})\\
L_{\geq 2}&=&\lambda !u.\lambda t.t!u
\end{eqnarray*}
This completes the proof.
\end{proof} 
}
{
Exploiting the classical strength of the \qcalc, we can easily
write terms computing the functions involved in the definition
of a finitely generated quantum circuit family. As a consequence, we
get the following result:}
\begin{theorem}\label{theo:circ-calc}
  For every finitely generated family of circuits $(f,g,h)$
  there is a quantum relevant term $M$ such that for each list
  $![!c_1,...,!c_n]$, $[1, M![!c_1,...,!c_n]]\Downarrow [Q,N]$
  (where $N= [r_1, \ldots, r_{m}]$) iff $m=f(n)$ and
  $Q=U_{\mathbf{K}_{h(k)}}(|r_1 \mapsto
  c_{g(n,1)},\ldots,r_{f(n)} \mapsto c_{g(n,f(n))}>)$
  (where we assume $c_0=0$ and $c_{n+1}=1$).
\end{theorem}

\subsection{From \textsf{Q}--calculus to Circuits}\label{subsect:qcal-circ}

We prove here the converse of theorem~\ref{theo:circ-calc}.
This way we will complete the proof of  the equivalence with quantum circuit families.

Let $M$ be a q--rel term, let $![!c_1,...,!c_n],
![!d_1,...,!d_n]$ be two lists of bits (with the same
length) and suppose\\
$[1, M![!c_1,...,!c_n]]\Downarrow_{\nqrul} [Q, N]$.
By applying exactly the same computation steps that
lead from $[1, M![!c_1,...,!c_n]]$ to $[Q, N]$, we can
prove that $[1, M![!d_1,...,!d_n]]\Downarrow_{\nqrul}[Q',N]$,
where $Q$ and $Q'$ live in the same Hilbert Space
$\Hs{\Qvt{N}}$.
Therefore, by means of  Church's Thesis, we obtain the following:
\begin{proposition}\label{prop:costruz-circ-I-fase}
  For each q--rel  $M$  there exist a term
  $N$ and two total computable
  functions $f:\NN \to \NN$ and 
  $g:\NN \times \NN \to \NN$ such that
  $[1, M![!c_1,...,!c_k]\Downarrow_{\nqrul} [|r_1 \mapsto
  c_{g(k,1)},\ldots,r_{f(k)} \mapsto c_{g(k,f(k))}>, N]$,
  where we conventionally set $c_0=0$ and $c_{n+1}=1$.
\end{proposition}

Let us consider $[\Qr,M]\in \Eqt$ and let us suppose that
$[\Qr,M]\Downarrow_{\quantrul} [\Qr', [r_1,\ldots,r_m]]$.  
The sequence of reductions in this
computation allows to to build in an effective way a unitary transformation 
$U_M$ such that $\Qr'= U_M(\Qr)$. 
Summarizing, we have the following:
\begin{proposition}\label{prop:costruz-circ-II-fase}
  Let $[\Qr,M]\in \Eqt$ and suppose
  $[\Qr,M]\Downarrow_{\quantrul}[\Qr',M']$. Then there is a circuit
  $K$ such that $\Qr' = U_K(\Qr)$. Moreover, $K$ is generated by
  gates appearing in $M$. Furthermore $K$ is effectively generated 
  from $M$.
\end{proposition}

As a direct consequence of propositions~\ref{prop:costruz-circ-I-fase}
and~\ref{prop:costruz-circ-II-fase} we obtain the following:

\begin{theorem}
  For each q--rel $M$ there is a quantum circuit family $(f,g,h)$ 
  such that for each list $![!c_1,...,!c_n]$,\\ if 
  $[1, M![!c_1,...,!c_n]]\Downarrow [\Qr,[r_1,\ldots,r_m]]$ then $m=f(n)$ and
  $\Qr=U_{\mathbf{K}_{h(n)}}(|r_1 \mapsto
  c_{g(n,1)},\ldots,r_{f(n)} \mapsto c_{g(n,f(n))}>)$.
\end{theorem}

\section{On the Measurement Operator}\label{sect:measure}
In the \qcalc\ it is not possible to classically observe the content
of the quantum register. More specifically, the language of terms does
not include any measurement operator which, applied to a quantum variable,
has the effect of observing the value of the related qubit. This is in contrast
with Selinger and Valiron's $\lambda_{sv}$ (where such a measurement operator
is indeed part of the language of terms) and with other calculi for quantum
computation like the so-called measurement calculus~\cite{Danos} (where the
possibility of observing is even more central). 

Extending \qcalc\ with a measurement operator $\mathtt{meas}(\cdot)$ (in the style 
of $\lambda_{sv}$) would not be particularly problematic. However, some of
the properties we proved here would not be true anymore. In particular:
\begin{varitemize}
\item
  The reduction relation would be probabilistic, since observing a qubit
  can have different outcomes. As a consequence, confluence would not be
  true anymore.
\item
  The standardization theorem would not hold in the form it has been presented
  here. In particular, the application of unitary transformations to the
  underlying quantum register could not always be postponed at the end of
  a computation.
\end{varitemize}

The main reason why we have focused our attention to a calculus without
any explicit measurement operator is that the (extensional)
expressive power of the obtained calculus would presumably be the
same~\cite{NieCh00}.

\section{Conclusion and Further Work}
We have studied the \qcalc, a quantum lambda calculus based on the
paradigm ``quantum data and classical control''. 
Differently from most of the related literature, which focus on semantical
issues, we faced the problem of expressiveness, proving
the computational equivalence of our calculus with a suitable class of
quantum circuit families (or equivalently, with the Quantum
Turing Machines \`a la   Bernstein and Vazirani). 

We have also given a standardization theorem, that should help clarifying
the interaction between the classical and the quantum world (at least
in a $\lambda$--calculus setting). Syntactical properties of the calculus, 
such as subject reduction and confluence, have been studied.

The next step of our research will concern the development of type
systems.  An interesting question is the following:
is it possible to give type systems controlling  the (quantum) computational 
complexity of representable functions? 

\bibliographystyle{latex8}
\bibliography{biblio}

\end{document}